\documentclass[sigconf]{acmart}

\usepackage{amsmath,amsfonts}
 
\usepackage{amssymb}

\usepackage{wrapfig} % 浮动表格
\usepackage{pifont}
\usepackage{amssymb}
\usepackage{bbding} %对勾叉号

\usepackage{pifont} %use the command \ding

\usepackage{url}
\usepackage{hyperref}
\hypersetup{
    colorlinks=true,
    linkcolor=blue,
    filecolor=magenta,      
    urlcolor=magenta,
    citecolor = magenta,
    pdftitle={An Example},
    pdfpagemode=FullScreen,
    }
\usepackage{tikz}
\usetikzlibrary{arrows.meta}
\usetikzlibrary{automata,positioning}

\usepackage{threeparttable}
\usepackage{capt-of}% or \usepackage{caption}

\definecolor{tabred}{RGB}{230,36,0}%
\definecolor{tabgreen}{RGB}{0,116,21}%
\definecolor{taborange}{RGB}{250,124,30}%
\definecolor{tabbrown}{RGB}{171,70,0}%
\definecolor{tabyellow}{RGB}{251,253,169}%
\newcommand*{\vcorr}{%
  \vadjust{\vspace{-\dp\csname @arstrutbox\endcsname}}%
  \global\let\vcorr\relax
}% 

\usepackage{mathtools}

\usepackage{lscape}
\usepackage[figuresright]{rotating}
\usepackage[graphicx]{realboxes}
\usepackage{adjustbox}
\usepackage{caption}

\usepackage{subfigure}

\usepackage{colortbl} 
\usepackage{xfrac}

\usepackage{appendix}
\usepackage{float}

\usepackage{fbox} %use box
\usepackage{fancybox}%use other types of boxes
\usepackage{framed}

\usepackage{multirow}

\def\BibTeX{{\rm B\kern-.05em{\sc i\kern-.025em b}\kern-.08em
    T\kern-.1667em\lower.7ex\hbox{E}\kern-.125emX}}
    
\usepackage{CJKutf8}  
\usepackage{pgfplots}
\usepackage{filecontents}

\usepackage{hyperref}

\usepackage{tabularx,makecell, caption}
\usepackage{enumitem} % to control Itemization spacing

\newcolumntype{L}{>{\arraybackslash}X}

\usepackage{multicol}
\usepackage{listings}
\usepackage{blindtext}
\usepackage{etoolbox,xstring,mfirstuc,textcase}
\usepackage{listings}

\usepackage{subfiles}
\usepackage[most]{tcolorbox}

\usepackage{xcolor}
\theoremstyle{plain}                
\theoremstyle{definition}       

\usepackage{siunitx}
\usepackage{comment}
\usepackage{indentfirst} 
\usepackage{framed} 

\usepackage[font=small,labelfont=bf,tableposition=top]{caption}
\usepackage{booktabs}
\usepackage{dashbox}

\usepackage{listings}
\usepackage{color}
\usepackage{algorithm,algpseudocode} %% select one of them

\renewcommand\footnotetextcopyrightpermission[1]{} % removes footnote with conference information in first column
\begin{document}
\title{Cryptocurrency in the Aftermath: Unveiling the \\ Impact of the SVB Collapse}

%=================================================
%author
%=================================================

%\begin{comment}  % priority

\author{Qin Wang$^{*}$, Guangsheng Yu$^*$, Shiping Chen$^*$}
\vspace{0.15in}
\affiliation{
\textit{$^*$CSIRO Data61, Australia} 
}

%\end{comment}

%=================================================
%abstract
%=================================================

\begin{abstract}
%The collapse of Silicon Valley Bank (SVB) in early 2023 sent shockwaves throughout financial markets. In this paper, we demystify the aftermath specifically on crypto markets. By undertaking an extensive four-month investigation encompassing over 100 mainstream cryptocurrencies, we unveil an array of compelling facts.

In this paper, we explore the aftermath of the Silicon Valley Bank (SVB) collapse, with a particular focus on its impact on crypto markets. We conduct a multi-dimensional investigation, which includes a factual summary, analysis of user sentiment, and examination of market performance. Based on such efforts, we uncover a somewhat counterintuitive finding: \textit{the SVB collapse did not lead to the destruction of cryptocurrencies; instead, they displayed resilience.} 

\end{abstract}

%\keywords{Cryptocurrency, SVB Collapse, HCI, Economics}

%=================================================
\maketitle

%=================================================   
%=================================================   
\section{Introduction}
\label{sec-intro}
%=================================================   

Silicon Valley Bank (SVB) met an unprecedented downfall in early 2023 owing to a severe bank run, ranking as the third-largest bank failure in the history of the U.S. since the 2007-2008 financial crisis. The far-reaching consequences have not spared the crypto markets. Comprehending the impact on such an emerging asset class becomes imperative. We retrace the entire story and dive into this crucial yet overlooked interplay. 

\smallskip
\noindent\textbf{SVB's sudden collapse}~\cite{timeline,nytnews}.
On \textit{March 8}, SVB reported a big loss of \$1.8 billion due to falling bond values caused by aggressive interest rate hikes. SVB planned to raise \$2 billion to stabilize its finances. On \textit{March 9}, SVB's stock price plummeted by 60\%, causing worries about its financial health. Depositors, mainly venture capital firms and tech startups, started withdrawing funds, including the Founders Fund, led by Peter Thiel. On \textit{March 10}, SVB faced a crisis as funds were withdrawn rapidly, making it almost collapse due to cash shortages. The Federal Deposit Insurance Corporation (FDIC) stepped in to protect depositors, ensuring up to \$250,000 per account, even though some had more than that. On \textit{March 11}, influential investors and tech leaders criticized the government for not doing more to help beyond the \$250,000 deposit guarantee. Billionaire Bill Ackman voiced his concerns on Twitter. On \textit{March 12}, Signature Bank in New York shut down due to fears of a bank run, especially among risky asset holders. In response, the FDIC, Treasury Department, and Federal Reserve assured depositors of both SVB and Signature Bank that their funds, even exceeding \$250,000, would be safe. On \textit{March 13}, President Biden tried to reassure the public about the crisis's impact on the financial system, but many bank stocks dropped significantly. First Republic Bank fell 65\%, and Charles Schwab, a major U.S. bank, dropped 11\%. On \textit{March 14}, The Federal Reserve Board announced a review of SVB's oversight due to its collapse. Additionally, investigations by the Justice Department and the Securities and Exchange Commission (SEC) into SVB's fall were reported by The Wall Street Journal.

\smallskip
\noindent\textbf{Ripple effect \& banking crisis.}
SVB provided banking services and credit to thousands of startup companies. However, the collapse has triggered a chain reaction of adverse consequences~\cite{bankcrisis}, including debt defaults and significant losses in depositor assets. These repercussions rippled across various sectors. Among the most severely affected were startups, constituting a substantial portion of SVB's client base. The collapse thrust them into a liquidity crisis, curtailing their access to essential capital. It not only imperiled the day-to-day operations but also sent shockwaves throughout the wider venture capital ecosystem. Beyond startups, the banking industry faced a multifaceted set of challenges (cf. \textcolor{teal}{Tab.\ref{tab-timeline}}), including the potential for collapses or mergers and acquisitions, ranging from regional institutions (e.g., Silvergate,  Signature) to traditional banks (e.g., First Republic, Credit Suisse). While the specific drivers might exhibit differences, SVB's collapse acted at least as a catalyst. Moreover, the reverberations extended into broader financial markets. They eroded market confidence and raised the specter of a financial crisis on the horizon.

\vspace{-0.05in}
\begin{table}[ht]
\caption{A timeline of the fallout in early 2023}
\vspace{-0.1in}
\label{tab-timeline}
\centering
\begin{threeparttable}
\resizebox{1\linewidth}{!}{
\begin{tabular}{c|c|c|c|c|c}
%\toprule
\multicolumn{1}{c}{\rotatebox{0}{\textbf{\textit{\makecell{Time}}}}} & 
\multicolumn{1}{c}{\rotatebox{0}{\cellcolor{gray!10}\textbf{\textit{\makecell{Bank}}}}} & 
\multicolumn{1}{c}{\rotatebox{0}{\cellcolor{gray!10}\textbf{\textit{\makecell{Customer}}}}} & 
\multicolumn{1}{c}{\rotatebox{0}{\cellcolor{gray!10}\textbf{\textit{Lo.}}}} & 
\multicolumn{1}{c}{\rotatebox{0}{\cellcolor{gray!10}\textbf{\textit{\makecell{Size}}}}} & 
\multicolumn{1}{c}{\rotatebox{0}{\cellcolor{gray!10}\textbf{\textit{\makecell{Event}}}}} 
\\ 
 
\cline{2-6}
\cellcolor{gray!10}Mar.  9 & \cellcolor{gray!10}Silvergate B.  & \cellcolor{gray!10}Crypto startups & \cellcolor{gray!10} US & \cellcolor{gray!10}\$11.35b     & \cellcolor{gray!10}Collapse \\
\cellcolor{gray!10}Mar. 10 & \cellcolor{gray!10}Silicon Valley B. &\cellcolor{gray!10} Hitech/Crypto & \cellcolor{gray!10}US  & \cellcolor{gray!10}\$209b &  \cellcolor{gray!10}Collapse \\
\cellcolor{gray!10}Mar.  12 & \cellcolor{gray!10}Signature B.  & \cellcolor{gray!10}Crypto startups & \cellcolor{gray!10}US &   \cellcolor{gray!10} \$110b   & \cellcolor{gray!10}Collapse \\

\cellcolor{gray!10}Mar. 19 & \cellcolor{gray!10}Credit Suisse   &\cellcolor{gray!10}  Public (global) & \cellcolor{gray!10}EU & \cellcolor{gray!10} \$1.5t &   \cellcolor{gray!10}Acquisition (UBS) \\ 

%\cellcolor{gray!10}Apr. & \cellcolor{gray!10}MUFG Union B.  & \cellcolor{gray!10}General & \cellcolor{gray!10}EU & \cellcolor{gray!10}\$163b  &  \cellcolor{gray!10}Close \\

\cellcolor{gray!10} Mar. - & \cellcolor{gray!10}Western Alliance B. &\cellcolor{gray!10}Public (regional)  & \cellcolor{gray!10}US &  \cellcolor{gray!10}\$67.7b  &  \cellcolor{gray!10} Share price fell 82\% \\

\cellcolor{gray!10}Apr. 28 & \cellcolor{gray!10}First Republic B.  &\cellcolor{gray!10}Public (regional)  & \cellcolor{gray!10}US &  \cellcolor{gray!10}\$271b  &  \cellcolor{gray!10}Acquisition (JP Morgan) \\

\cellcolor{gray!10}  Jul. 25 & \cellcolor{gray!10}PacWest B. &\cellcolor{gray!10}Public (regional)  & \cellcolor{gray!10}US &  \cellcolor{gray!10}\$1.4b  &  \cellcolor{gray!10} Acquisition (B. of California)  \\

\end{tabular}
}
\begin{tablenotes} 
       \footnotesize
       \item[] \textit{Fund size} is recorded on Dec. 2022 and measured in US dollars.
\end{tablenotes}
\end{threeparttable}
\end{table}
\vspace{-0.1in}

\smallskip
\noindent\textbf{A systemic risk or just contagion?} After a series of bank failures, the Federal Reserve implemented emergency measures (as previously mentioned) to provide support to banks facing short-term liquidity challenges and prevent bank runs akin to the crises witnessed at SVB and Signature. The debate emerged whether SVB's collapse would be confined to a contagion within the banking system or evolve into a systemic risk affecting the broader financial system. Over several months (from March to September), intuitive observations indicated a gradual stabilization of the banking system, with the initial impact diminishing. A perspective outlined in~\cite{oxford2023svb} suggested that SVB's failure contributed to contagion but did not pose a systemic risk. However, it is important to note that while competitive banks managed to recover their losses,  the plight of customers who deposited significant sums in the failed banks has been overlooked. In this context, our focus shifts to crypto-companies, a substantial portion of SVB's depositors.

\smallskip
\noindent\textbf{In quest of attention towards cryptoassets.} Analogous to conventional public banks that hold fiat currencies as the foundation of entire economies, the crypto/Web3 space relies on stablecoins as a parallel anchor, symbolizing a mirrored iteration of fiat money. These stablecoins, such as USDT, are designed to maintain a fixed exchange rate with their fiat counterparts, like the USD, at a 1:1 ratio. Crypto companies operating in this space often securitize real-world assets (RWA) and issue on-chain tokens, facilitating their circulation within the blockchain ecosystem. Notably, these RWA assets have commonly been deposited with SVB. Based on available resources~\cite{Cointelegraph,decrypt,forbes}, SVB's client roster includes several prominent crypto industry players, including Ripple (with deposits of approximately \$1b), BlockFi (\$227m), Circle (\$3.3b), Pantera (\$560m), Avalanche (\$1.6m), Roku (\$487m), Roblox (\$150m), Ginkgo Bioworks Holdings (\$74m), iRhythm Technologies (\$54.5m), Rocket Lab (\$38m), LendingClub (\$21m),  Yuga Labs (in limited capacity), Proof (to some extent), Nova Labs (to some extent), and Dapper Labs (to a minimal extent). Concurrently, Signature Bank's clientele includes Coinbase (\$240m), Celsius, and Paxos (\$260m). While these institutions do not represent the entirety of the crypto market, the potential erosion of reserves within any crypto project holds significant implications.

Furthermore, the absence of comprehensive research has fueled debates within communities. Advocates contend that the SVB crisis underscores the essential nature of decentralized and transparent financial systems exemplified by blockchain technology. However, critics highlight the immediate ripple effects experienced within the crypto realm, particularly the temporary de-pegging of the stablecoin USDC, as compelling evidence that crypto assets are not immune to systemic financial risks. This episode has also catalyzed intensified regulatory scrutiny, accompanied by allegations that the crypto sector is unjustly blamed for broader systemic failures within the financial system. Amidst this intricate landscape, the pressing need for decent research into the impact of the SVB collapse on cryptocurrencies becomes increasingly evident. This motivates our exploration of this critical subject matter.

%Guided by the central question, \textit{What are the impacts on the crypto space post-SVB's collapse: any factual developments, speculative narratives, or interesting findings?}

\smallskip
\noindent\textbf{Our efforts} (\textcolor{teal}{Sec.\ref{sec-metho}}).  We advance the exploration as follows.

\smallskip
\noindent \textcolor{teal}{\ding{172}} \textit{\textbf{We dive into SVB and the cause of its collapse}} (\textcolor{teal}{Sec.\ref{sec-svb}}). Despite SVB's successful 40-year track record of development, the collapse is not entirely surprising. The root reasons can be attributed to a sequence of high-risk financial decisions, both external and internal in nature (\textcolor{teal}{Sec.\ref{subsec-factors}}). In response to the liquidity injections and interest rate reductions initiated by the FED during COVID-19, SVB witnessed a nearly threefold surge in its deposits. However, to maximize returns, the bank made significant investments in long-term, fixed-rate government bonds and mortgage-backed securities. As the FED conducted an aggressive interest rate hike in late 2021, SVB's vulnerability to these rate increases became increasingly apparent. Owing to its imbalanced portfolio, a wave of withdrawals by its depositors ensued, resulting in substantial unrealized losses that had to be realized.

\smallskip
\noindent \textcolor{teal}{\ding{173}} \textit{\textbf{We dissect components in the crypto space and present their interconnections with traditional finances}} (\textcolor{teal}{Sec.\ref{sec-cryto}}). Specifically, our investigation covers three main aspects: (i) differentiating cryptoassets from traditional banks based on their features, namely, intermediaries, regulation, and reserve credit (\textcolor{teal}{Sec.\ref{subsec-tybank}});
(ii) classifying cryptoassets by their nature and usage, which includes platform tokens, X-20 tokens, stablecoins, and derivatives (\textcolor{teal}{Sec.\ref{subsec-classify}}); and (iii) examining the technical underpinnings, with a particular focus on understanding the decoupled blockchain layers (\textcolor{teal}{Sec.\ref{subsec-decouple}}).

\smallskip
\noindent \textcolor{teal}{\ding{174}} \textit{\textbf{We undertake an extensive empirical study}} (\textcolor{teal}{Sec.\ref{sec-metho}\textcolor{black}{\small\&}Sec.\ref{sec-emprical}}), to comprehensively explore the relationship between cryptoassets and the SVB collapse event. Our investigation encompasses multiple dimensions, including: (i) analyzing the price trends of nine prominent cryptoassets (\textcolor{teal}{Sec.\ref{subsec-crypto-price}});
(ii) evaluating sentiment trends through the study of tweets (\textcolor{teal}{Sec.\ref{subsec-sentiment-tweet}}\&\textcolor{teal}{Sec.\ref{subsec-senti-composite}}); and (iii) examining market behaviors using a series of technical indicators (\textcolor{teal}{Sec.\ref{subsec-market-behav}}).

\smallskip
\noindent \textcolor{teal}{\ding{175}} \textit{\textbf{We offer comprehensive factual explorations}} (\textcolor{teal}{Sec.\ref{sec-history}}) of significant events in crypto history, including major advancements (\textcolor{teal}{Sec.\ref{subsec-tokenwave}}) and human-based failures (\textcolor{teal}{Sec.\ref{subsec-fail}}). Additionally, \textit{\textbf{we engage in thoughtful discussions}} regarding the impact of the collapse (\textcolor{teal}{Sec.\ref{sec-impac}}), provide  \textit{\textbf{summaries of user perceptions}}(\textcolor{teal}{Sec.\ref{sec-userpcpt}}), and \textit{\textbf{point out potential new challenges}} (\textcolor{teal}{Sec.\ref{sec-chall}}). To enhance understanding, we also furnish supplementary knowledge on both traditional financial systems (a.k.a., centralized finance or CeFi) and decentralized finance (DeFi) in \textcolor{teal}{Appendix A}\&\textcolor{teal}{B}.

\smallskip
\noindent \textcolor{teal}{\ding{171}} \textbf{Our findings.} Through our investigation, we uncover a series of undiscovered facts that highlight the impact of the SVB collapse on the crypto markets. Some of the key findings include:

\begin{comment}

Key takeaways
\begin{itemize}
    \item All the cryptocurrency prices are closely relevant. 
    \item Cryptocurrency assets distributions in traditional banks (xxx\%).  
    \item The dropdown cryptocurrency prices are aligned with the primary stock market (xxx\%).  
    \item Cryptocurrency stablecoins are slightly de-peg (up to xxx\%) but recovers quickly (within xx days). 
    \begin{itemize}
    \item Non-custodian:
    \item Custodian by fiat money:
    \item Custodian by cryptocurrencies:
    \end{itemize}
    \item Long term low prices 
    \item 2-8 resistance
\end{itemize}

\end{comment}

\smallskip
\noindent\textcolor{teal}{\ding{49}} \textit{\textbf{Root cause recap}}. The downfall of SVB can be attributed to a combination of external and internal factors. Externally, the macroeconomic environment was influenced by the impact of COVID-19 and consecutive rate hikes by the Federal Reserve, which altered market expectations negatively (\textcolor{teal}{\textbf{Observation-3.\ding{204}}}). Internally, SVB's reliance on a singular asset management strategy  (\textcolor{teal}{\textbf{Observation-3.\ding{203}}}), primarily centered on long-term fixed bonds, led to a lack of flexibility and liquidity. Additionally, the skewed portfolio, heavily weighted toward high-tech startups with unstable profits, exacerbated the downward spiral (\textcolor{teal}{\textbf{Observation-3.\ding{205}}}).

\smallskip
\noindent\textcolor{teal}{\ding{49}} \textit{\textbf{Correlation patterns}}. We observe explicit correlation patterns with a high level of coherence: (i) among different cryptocurrencies within the same type (\textcolor{teal}{\textbf{Observation-5.\ding{202}}}\&\textcolor{teal}{\textbf{5.\ding{203}}}), (ii) between user sentiment and asset prices  (\textcolor{teal}{Fig.\ref{fig:tweets}}), (iii) between asset market performance and macroeconomic trends in traditional finance (\textcolor{teal}{\textbf{Observation-5.\ding{207}}}). Additionally, we anticipate an imminent convergence of CeFi and DeFi and an increasing embrace of government-based regulatory measures in the crypto space (\textcolor{teal}{Sec.\ref{sec-chall}}).

\smallskip
\noindent\textcolor{teal}{\ding{49}}  \textit{\textbf{Market volatility}}. We observed that the crypto markets experienced significant fluctuations in volatility following the SVB collapse, reflecting increased market uncertainty. Somewhat surprisingly, the behavior of cryptoassets performed not as poorly as expected (either token's pricing dropping or stablecoin's deppegging). SVB's fallout merely contributed to short-term market volatility (compared to the period before collapse rather than its peak time), and the crypto markets demonstrated overall resilience in the face of such a significant failure (\textcolor{teal}{\textbf{Observation-5.\ding{202}}}\&\textcolor{teal}{\textbf{5.\ding{203}}}\&\textcolor{teal}{\textbf{5.\ding{207}}}).

\smallskip
\noindent\textcolor{teal}{\ding{49}}  \textit{\textbf{Investor sentiment}}. Investor attention has significantly increased (\textcolor{teal}{\textbf{Observation-5.\ding{203}}}) following the collapse of SVB, but their explicit sentiments are complex, ranging from optimism to pessimism (\textcolor{teal}{Sec.\ref{subsec-user-coexist}}). Fortunately, based on the available data, albeit limited, the prevalence of positive sentiment outweighs the negative (\textcolor{teal}{\textbf{Observation-5.\ding{204}}}\&\textcolor{teal}{\textbf{5.\ding{205}}}). Moreover, the collapse may have influenced investors' activities in the crypto markets, potentially leading to changes in risk appetite and portfolio strategies (\textcolor{teal}{Sec.\ref{subsec-user-lesson}}).

\smallskip
\noindent\textcolor{teal}{\ding{49}}  \textit{\textbf{Asset class differentiation}}. While cryptocurrencies as a whole exhibit a high degree of coherence, differentiation exists among specific assets, a phenomenon applicable to all types of crypto assets. During the fallout, certain assets demonstrated strong resilience, while others faced substantial flunctions (\textcolor{teal}{Fig.\ref{fig:closing_price}}). We infer that this phenomenon follows a \textit{2-8 effect}, where the more well-known and highly ranked an asset is, the more stable and resilient it tends to be, and the faster it recovers to an expected level.

%=================================================   
\section{Methodology}
\label{sec-metho}
%===============================================

\subsection{Pilot Study}

\noindent\textbf{Existing resources.} Vo et al.~\cite{vo2023hero} explore the collapse of Silicon Valley Bank, revealing a series of missteps that led to its failure spanned from 2019-2022. Key issues include risky investments in debt securities amid low-interest rates, a lack of depositor diversification, and poor risk management infrastructure.  Dewatripont et al.~\cite{dewatripont2023silicon} provide a study for prudential regulation both in Europe and globally from the SVB collapse. The study underscores the urgency for the European Union to refine its bank resolution frameworks, particularly concerning the rigid \textit{8\% bail-in rule} of the Banking Recovery and Resolution Directive (BRRD). Additionally, the collapse exposes flaws in the current treatment of short-term corporate deposits, calling for a reevaluation of insurance caps and liquidity requirements. Aharon et al.~\cite{aharon2023too} investigated the ripple effects of the SVB collapse on global equity markets, finding a largely negative response across Europe, Latin America, the Middle East, and Africa. Another finding is the Asian market showed a delayed but significant negative reaction in the days following the event. More analyses of the SVB collapse to traditional (macro-)economics and finances refer to~\cite{yousaf2023impact,pandey2023repercussions,akhtaruzzaman2023did,ciuriak2023silicon,perdichizzi2023non,benmelech2023bank,bales2023public}.

In contrast, research on cryptocurrencies is lacking, with only a limited number of studies providing primary analyses. Mustafa et al.~\cite{svb2cryto} conducted an investigation into the impact of SVB on stablecoins. Their study highlights the event where USDC experienced devaluation, losing its 1:1 peg with the U.S. dollar. This devaluation was a consequence of Circle, the issuer of USDC, having 8\% of its reserves—equivalent to \$3.3 billion—deposited in SVB. As a result, users rushed to swap their USDC for other stablecoins, causing a surge in transaction fees and driving up gas prices on the Ethereum network.  Galati~\cite{galati4488220silicon} examined the contagion effects by using a BEKK model to investigate the impact of the SVB bankruptcy on stablecoins, a market that was significantly affected by large abnormal movements and inefficiency driven by traders' behavior. The study's uniqueness lies in its use of proprietary min-by-min data of major cryptoassets and its analysis of a singular exogenous event.

\smallskip
\noindent\textbf{Identified gaps.} We highlight several gaps for the existing body of research on the SVB collapse.

\begin{itemize}
    \item \textit{Lack of focus on crypto:} While many studies analyze the repercussions of SVB's collapse on a broad scale, including its impact on global equity markets, prudential regulation, and financial stability, there is a deficiency in research that dive into its specific effects on crypto markets.

    \item \textit{Fragmented analysis of RWA with cryptoassets:} The studies mentioned in this context primarily center on equity markets, risk management, and regulatory frameworks. However, they leave unanswered questions regarding how crypto markets responded to this shock and the intricate relationship between centralized finance and decentralized finance.

    \item\textit{Limited multi-angle examination:} Existing studies restrict focus to a narrow subset of cryptoassets, particularly stablecoins, while neglecting major cryptocurrencies such as Bitcoin or Ethereum. Additionally, these studies overlook aspects such as the impact on investor sentiment, user behaviors, and market volatility.

    \item \textit{Superficial analysis:} Many studies provide surface-level explanations and insights derived from similar sources, lacking in-depth exploration of more advanced materials.
    
    \item \textit{Absence of systematic insights:} Most public resources describe the situation without delving into the valuable lessons we can extract from this collapse.
\end{itemize}

\smallskip
\noindent\textbf{Our motivation.}
Motivated by these research gaps, our study centers around the research question (\textit{\textbf{RQ.0}}) \textit{what are the impacts on the crypto space post-SVB's collapse: any factual developments, speculative narratives, or interesting findings?} 
Following this overarching question, we dissect it into a series of sub-questions as follows:

\vspace{-0.1in}
\begin{table}[!hbt]
\caption{Research questions}
\label{tab-question}
\vspace{-0.13in}
\resizebox{\linewidth}{!}{
\begin{tabular}{c|l} 
\cellcolor{gray!10} \textit{\textbf{RQ.1}} & \cellcolor{gray!10}What is SVB, and any distinctive features?   \\ 

\cellcolor{gray!10} \textit{\textbf{RQ.2}} &\cellcolor{gray!10} What are the procedure and root causes of SVB's collapse? \\ 

 \cellcolor{gray!10}  \textit{\textbf{RQ.3}} & \cellcolor{gray!10} What are cryptoassets and key characteristics? \\ 

 \cellcolor{gray!10} \textit{\textbf{RQ.4}} & \cellcolor{gray!10} How do cryptoassets differ from conventional finance? \\ 

\cellcolor{gray!10}  \textit{\textbf{RQ.5}} & \cellcolor{gray!10} What transpired in the crypto space after SVB's downfall?  \\ 

 \cellcolor{gray!10}  \textit{\textbf{RQ.6}}  & \cellcolor{gray!10} Any shifts in user behavior, stakeholder involvement, or market dynamics?   \\ 

 \cellcolor{gray!10}  \textit{\textbf{RQ.7}} & \cellcolor{gray!10}  What impacts have on the crypto ecosystem?  \\ 

 \cellcolor{gray!10} \textit{\textbf{RQ.8}} & \cellcolor{gray!10}  What valuable insights can we glean? \\ 
\end{tabular}
}
\end{table}

%\begin{itemize}
%    \item[\textit{\textbf{RQ.1}}] What is SVB, and any distinctive features?
%    \item[\textit{\textbf{RQ.2}}] What are the procedure and root causes of SVB's collapse?
%    \item[\textit{\textbf{RQ.3}}] What are cryptoassets and key characteristics?
%    \item[\textit{\textbf{RQ.4}}] How do cryptoassets differ from conventional finance?
%    \item[\textit{\textbf{RQ.5}}] What transpired in the crypto space after SVB's downfall?
%    \item[\textit{\textbf{RQ.6}}] Were there significant shifts in user behavior, stakeholder involvement, or market dynamics?
%    \item[\textit{\textbf{RQ.7}}] What impacts have on the crypto ecosystem?
%    \item[\textit{\textbf{RQ.8}}] What valuable insights can we glean?
%\end{itemize}

\subsection{Methodology}

\begin{figure}[!hbt]
    \centering
    \includegraphics[width=\linewidth]{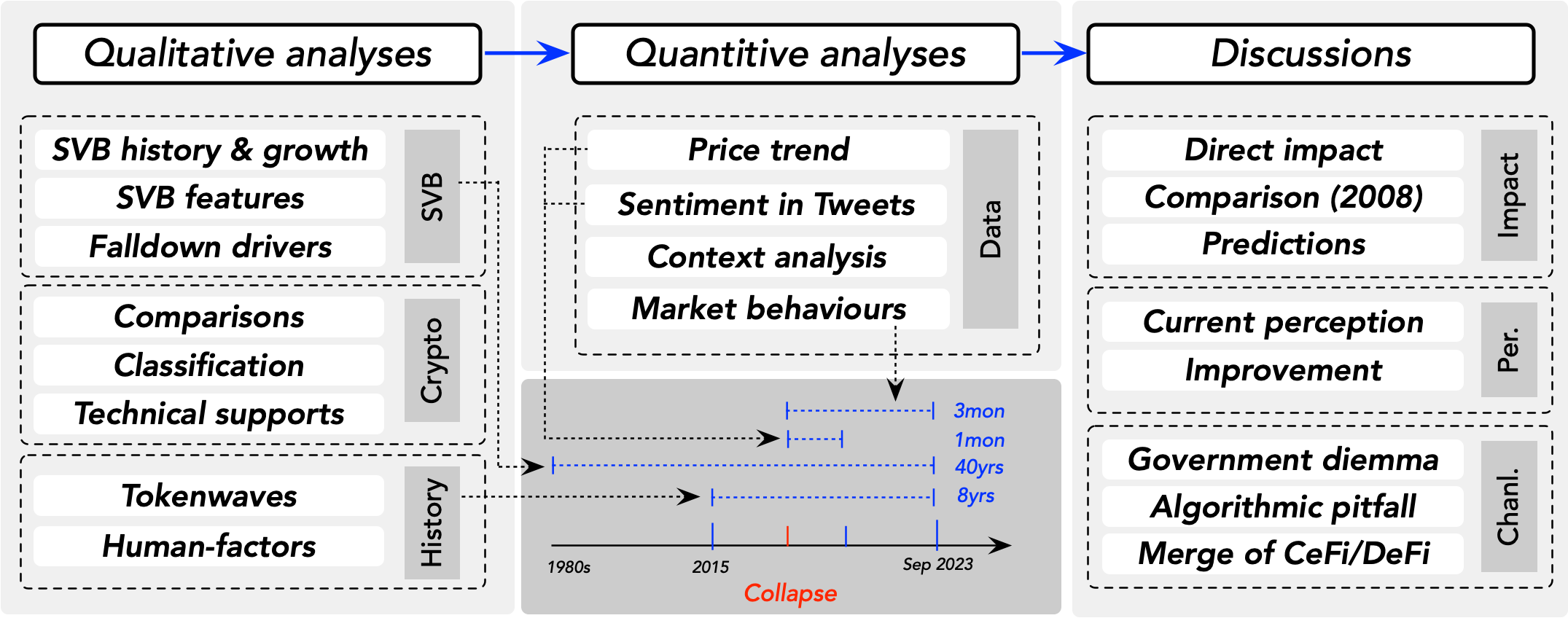}
    \vspace{-0.2in}
   \caption{Methodology}
   \label{fig-meth}
\end{figure}

For completeness, we employ a dual approach (\textcolor{teal}{Fig.\ref{fig-meth}}) to our exploration, combining qualitative and quantitative investigations.

\smallskip
\noindent\textbf{Qualitive analysis.} In this study, we first explore the fundamental knowledge of two entities, namely, SVB (\textcolor{teal}{Sec.\ref{sec-svb}}) and cryptoassets (\textcolor{teal}{Sec.\ref{sec-cryto}}). Combining this with a review of the long-term growth of SVB through public resources (e.g., news, reports, articles, blogs), we identify two major features (\textcolor{teal}{Sec.\ref{subsec-svb-feature}}) that can straightforwardly describe the operation of SVB and its composition. Driven by this, we investigate the driving factors  (\textcolor{teal}{Sec.\ref{subsec-factors}}) leading to its collapse.

    \vspace{-0.15in}
    
For cryptoassets, we examine their comparisons  (\textcolor{teal}{Sec.\ref{subsec-tybank}}) with traditional financial systems and provide an overview of the classification of existing cryptoassets (\textcolor{teal}{Sec.\ref{subsec-classify}}). We also introduce their technical support  (\textcolor{teal}{Sec.\ref{subsec-decouple}}), which is primarily based on blockchain and smart contracts. Beyond this, we provide an overview of historical token waves driven by innovations in the cryptoasset space  (\textcolor{teal}{Sec.\ref{subsec-tokenwave}}) and present the historical events of crypto failures  (\textcolor{teal}{Sec.\ref{subsec-fail}}). These elements collectively serve as a suitable benchmark.

\smallskip
\noindent\textbf{Quantitative analysis.} We explore data from various public sources across three distinct dimensions:

\begin{itemize}
    \item \textbf{Exploring SVB's growth} \textit{(spanning \textcolor{teal}{40 years}}, cf. \textcolor{teal}{Sec.\ref{subsec-svb-growth}}). In this dimension, our primary focus is on selectively studying data related to SVB over a period of 40 years, tracing its evolution since its inception. We aim to discern pivotal development stages, identifying key features such as major clientele and portfolio composition (as discussed before). We rely on multiple publicly available resources, including balance sheets and company composition records.
    \item \textbf{Crypto market analysis.} This dimension includes an examination of crypto markets (\textcolor{teal}{Sec.\ref{sec-emprical}}), consisting of two facets: 
        \begin{itemize}
       \item  \textit{Crypto asset assessment (covering \textcolor{teal}{one month} post-collapse}, cf. \textcolor{teal}{Sec.\ref{subsec-crypto-price}}). We analyze nine major crypto assets, categorizing them into three representative types. The data spans one month from the date of the market collapse (March).
       \item  \textit{Market behavior assessment (over \textcolor{teal}{three months}},  cf. \textcolor{teal}{Sec.\ref{subsec-market-behav}}). We delve into the market's behavior using three essential technical indicators: the Crypto Fear \& Greed Index, the Spent Output Profit Ratio, and the Pi Cycle Indicator. Our data analysis extends over a three-month period.
       \end{itemize}
    \item\textbf{Public sentiment analysis on Twitter} (\textit{over \textcolor{teal}{one month}}, cf. \textcolor{teal}{Sec.\ref{subsec-sentiment-tweet}}\&\textcolor{teal}{Sec.\ref{subsec-senti-composite}}). In this dimension, we gauge the sentiment of active crypto users and stakeholders on Twitter (renamed with X) over the course of one month. We collect data by monitoring relevant hashtags, such as \textit{\#crypto} and \textit{\#SVB/collapse} (also with several sub-tags). Our analysis includes the quantification of tweet volumes, assessing their significance, and conducting content analysis.
\end{itemize}

\noindent\textbf{Our answers.} Based on our analyses, we provide our results (\textcolor{teal}{Tab.\ref{tab-result}}) for each research question.

\vspace{-0.1in}
\begin{table}[!hbt]
\caption{Our results}
\label{tab-result}
\vspace{-0.15in}
\resizebox{\linewidth}{!}{
\begin{tabular}{c|ccc|} 

\multicolumn{2}{c}{\cellcolor{gray!10}\textbf{\text{Questions}}} &   \multicolumn{1}{c}{\cellcolor{gray!10}\textbf{\text{Ref.}}} &  \multicolumn{1}{c}{\cellcolor{gray!10} \rotatebox{0}{\textbf{\text{Key}}}} \\ 

\cmidrule{1-2}

\multirow{2}{*}{\rotatebox{90}{\textbf{SVB}}}  & \cellcolor{gray!10} \textit{\textbf{RQ.1}} & \cellcolor{gray!10}\textcolor{teal}{Sec.\ref{subsec-svb-growth}} \& \textcolor{teal}{Sec.\ref{subsec-svb-feature}} \ & \multicolumn{1}{c}{\makecell{\textit{rapid growth} with favor for \textit{hi-tech startups} \\ and holding \textit{long-term treasury bonds} } } \\ 

&  \cellcolor{gray!10} \textit{\textbf{RQ.2}} & \cellcolor{gray!10}  \textcolor{teal}{Sec.\ref{sec-intro}} \& \textcolor{teal}{Sec.\ref{subsec-factors}}  & \multicolumn{1}{c}{ \makecell{poor decisions on \textit{asset management} \\and tensed environment by \textit{rate hikes} } } \\ 

\cmidrule{1-2}

\multirow{2}{*}{\rotatebox{90}{\textbf{Cryp.}}}  & \cellcolor{gray!10}  \textit{\textbf{RQ.3}} & \cellcolor{gray!10} \cellcolor{gray!10}\textcolor{teal}{Sec.\ref{subsec-classify}}  & \multicolumn{1}{c}{ \makecell{operates \textit{on-chain} with decentralization,\\ mirrored \textit{DeFi products} compared to CeFi} } \\ 

&  \cellcolor{gray!10} \textit{\textbf{RQ.4}} & \cellcolor{gray!10} \textcolor{teal}{Sec.\ref{subsec-tybank}} \& \textcolor{teal}{Apdx.A}  \& \textcolor{teal}{Apdx.B}   &    \multicolumn{1}{c}{ \makecell{  \textit{loose} regulation, issurance \textit{without reserves}, \\ \textit{algorithmic} management, \textit{non}-intermediaries} } \\ 

\cmidrule{1-2}

\multirow{5}{*}{\rotatebox{90}{\textbf{Interplay}}}& \cellcolor{gray!10}  \textit{\textbf{RQ.5}} & \cellcolor{gray!10}  \cellcolor{gray!10}\textcolor{teal}{Sec.\ref{subsec-crypto-price}} \& \textcolor{teal}{Sec.\ref{sec-history}} &\multicolumn{1}{c}{  \makecell{ crypto-tokens \textit{price-drop} and \\ stablecoins \textit{deppeging}, but later \textit{recover} } }  \\ 

& \cellcolor{gray!10}  \textit{\textbf{RQ.6}} 
 & \cellcolor{gray!10}  \cellcolor{gray!10}\textcolor{teal}{Sec.\ref{subsec-crypto-price}} to \textcolor{teal}{Sec.\ref{subsec-market-behav}}  &  \makecell{ public attention \textit{increased} with  attitude\\ tending to be \textit{positive}, along with \textit{behaviors}  }    \\ 

& \cellcolor{gray!10}  \textit{\textbf{RQ.7}} & \cellcolor{gray!10}  \cellcolor{gray!10}\textcolor{teal}{Sec.\ref{sec-impac}} \& \textcolor{teal}{Sec.\ref{sec-userpcpt}}  &  \makecell{ \textit{regulation} and \textit{macroeconomic} affects \\ crypto, \textit{merging} DeFi \& CeFi is on road } \\ 

&  \cellcolor{gray!10} \textit{\textbf{RQ.8}} & \cellcolor{gray!10}   \textcolor{teal}{\textbf{Observation-3.\ding{202}}}  to  \textcolor{teal}{\textbf{6.\ding{203}}}  &  see \textcolor{teal}{Sec.\ref{sec-intro}}   \\ 

\cmidrule{4-4}

\end{tabular}
}
\end{table}

\subsection{Validity Discussion}

\noindent\textbf{Construct validity} This metric is used to indicate the degree to which the research questions and methodology are aptly employed in the study~\cite{chen1987theory}. The principal threat in our data analysis lies in the \textit{representativeness} of the public datasets used to understand the collapse of SVB and its impact on crypto markets. To mitigate this,  we carefully designed our data analysis techniques to minimize this bias. we have chosen datasets that span a range of metrics, including market performance, sentiment analysis, and regulatory responses. However, these datasets may still lack certain perspectives, such as those from internal SVB stakeholders or lesser-known crypto platforms. It remains possible that the data available in the public domain may emphasize issues that are more widely discussed.

\smallskip
\noindent\textbf{Internal validity.} It focuses on factors that may affect the validity of the results~\cite{bleijenbergh2011methodological}. The main threat lies in whether our research methodology and data analysis sufficiently answer the research questions. Our first weakness is the \textit{comprehensiveness} of the public datasets, which may influence the quality of research findings. The second weakness is that the data we collected may have a \textit{relatively short duration} due to numerous restrictions on accessing more valuable datasets. These limitations could impact the depth and applicability of our findings. To address these concerns, we have iteratively reviewed our data analysis and cross-validated our interpretations to ensure accuracy. We have also adopted the following steps to mitigate the weakness:

\vspace{-0.2em}
\begin{itemize}
\item In addition to examining data from websites and literature from academic resources, our analysis includes less commonly explored dimensions, drawing from academic journals, financial reports, and think-tank publications.

\item beyond quantitative analyses, our study also incorporates examinations of historical narratives, providing a foundational framework for comprehension and context.
\end{itemize}
\vspace{-0.2em}

\noindent\textbf{External validity.} The indicator pertains to the generalizability of our findings to different settings~\cite{calder1982concept}. Our results are largely based on public datasets and may not fully represent the complexities of the SVB collapse and its implications on the crypto industry. The public datasets are often biased towards highly visible events that may lead to limited applicability to broader domains. For mitigation, we provide subjective analyses and summarized insights following each section of the analysis, with the aim of improving its overall applicability and relevance.

% \qw{@sbr add limitations of Sec4 like short duration blabla}

% \textit{Short duration.}

% \textit{Limited cryptocurrency bucket.}

% \textit{Subjective induction.}

%==============================================
\section{Understanding SVB and its Collapse}
\label{sec-svb}
%==============================================
\begin{figure*}[t]
    \centering
    \subfigure[By country]{
    \begin{minipage}[t]{0.32\textwidth}
    \centering
    \includegraphics[width=\linewidth]{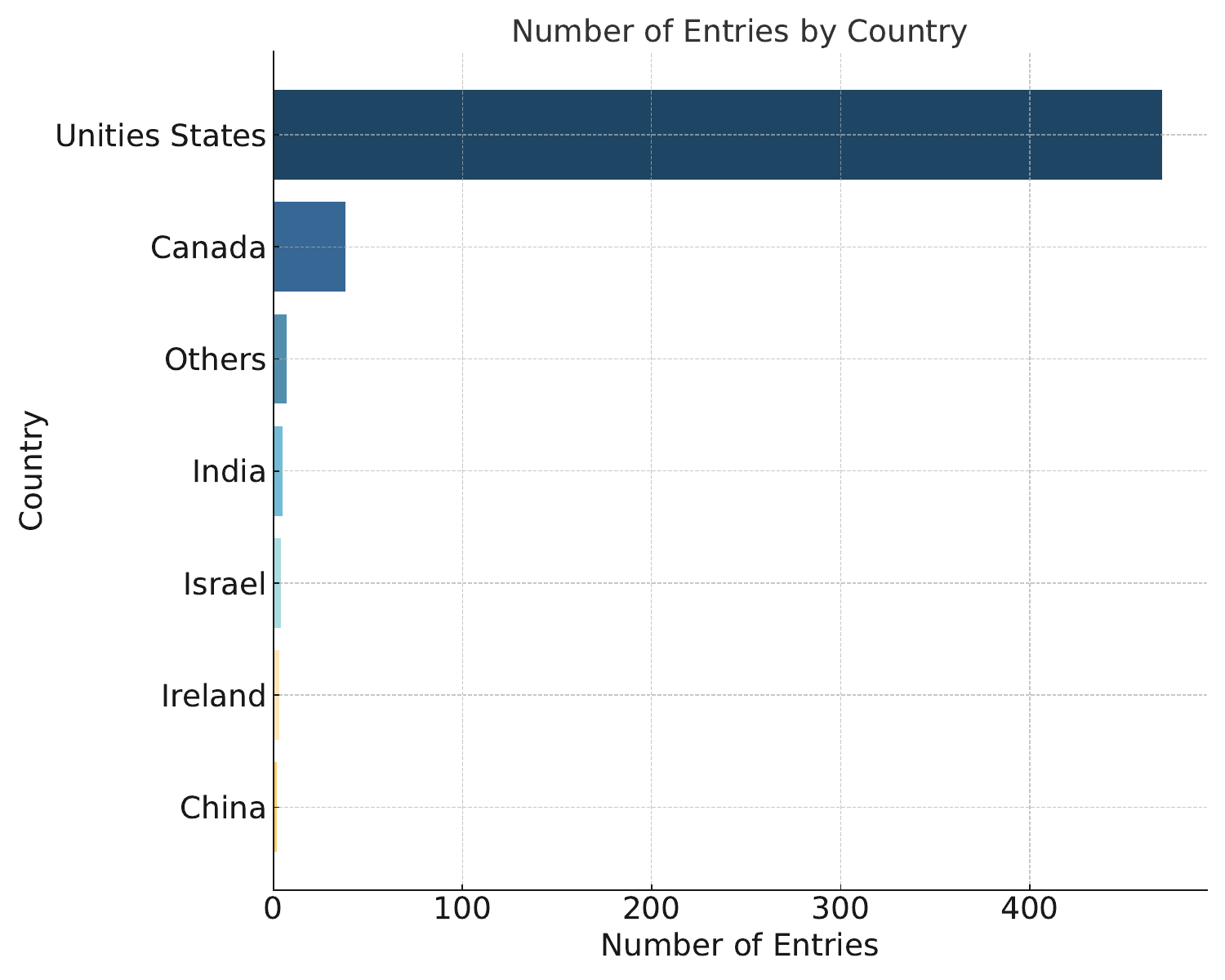}%2.4in
    \end{minipage}
    \label{fig:svb_portfolio_country}
    }
    \subfigure[By industry]{
    \begin{minipage}[t]{0.32\textwidth}
    \centering
    \includegraphics[width=\linewidth]{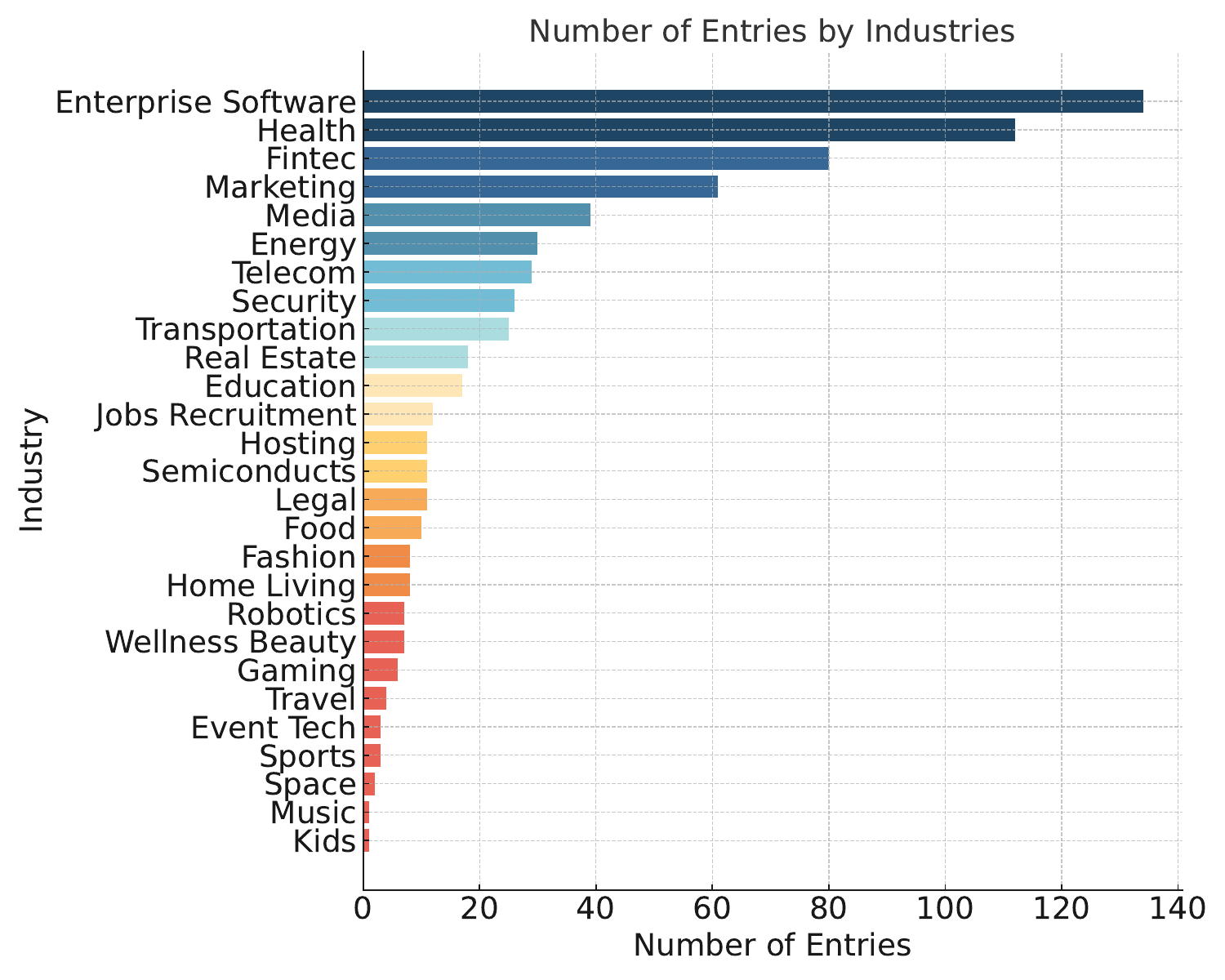}% 2.3in
    \end{minipage}
    \label{fig:svb_portfolio_industry}
    }
    \subfigure[By round]{
    \begin{minipage}[t]{0.32\textwidth}
    \centering
    \includegraphics[width=\linewidth]{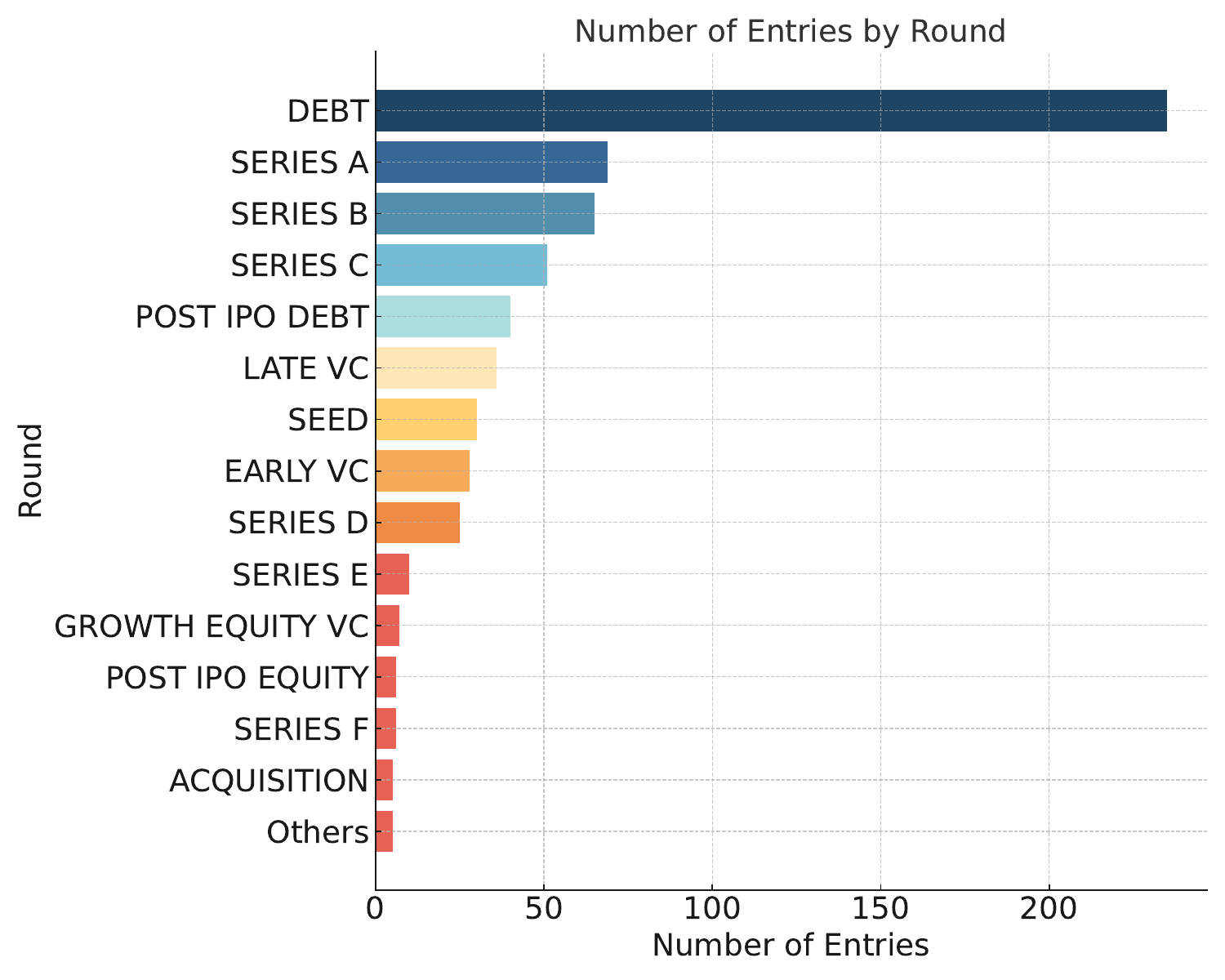}%2.3in
    \end{minipage}
 \label{fig:svb_portfolio_round}
    }
    \vspace{-0.15in}
    \caption{SVB portfolio analysis}
    \label{fig:svb_portfolio}
    \vspace{-0.15in}
\end{figure*}

\subsection{SVB History \& Growth}\label{subsec-svb-growth}
We briefly introduce a history recap of SVB (1980s-present) with analytical data sourced from its annual financial reports~\cite{svb2023,svb2023-2,svb2023-3,svb2023-4}.

\smallskip
\noindent\textbf{Late 1980s to early 1990.} SVB, founded in 1983, emerged during Silicon Valley's booming technological landscape. The founders recognized the opportunities offered by the region's venture capital environment and technological characteristics and launched a bank exclusively focused on serving technology startups. SVB experienced rapid growth and was listed on Nasdaq in 1988. By 1992, SVB's total assets reached \$960 million. The net profit per employee surpassed Citibank's by more than 4X in 1993. Over a decade, SVB successfully crafted its brand image, coinciding with the global popularity of the term \textit{Silicon Valley}.

\smallskip
\noindent\textbf{Early 1990s to early 2000s.} SVB also focuses on specific niche markets, particularly in assessing credit risks for small businesses during pre-lending, during-lending, and post-lending processes. Additionally, benefiting from relaxed geographical regulations, SVB expanded nationwide, achieving significant results from 1993 to 2001: assets increased from \$935 million to \$5.5 billion, the employee count exceeded 1000 from 235, and the market value peaked at over \$3 billion (from an initial public offering value of \$63 million). During this period, SVB's average shareholder return rate was an impressive 17.5\% per year, surpassing the 12.5\% average return rate of U.S. commercial banks.

\smallskip
\noindent\textbf{Early 2000s to late 2000s.}  After the internet bubble burst in 2001, SVB evolved into an integrated platform, offering equity investment, investment banking, asset management, and more to increase non-interest income. Concurrently, SVB strengthened risk control measures to mitigate systemic crises and capitalized on the trend of American venture capital funds expanding internationally. By 2004, SVB had established subsidiaries and consulting companies in various locations, internationalizing its operations in the UK, India, China, Israel, and others. By 2008, SVB fully recovered from the internet bubble crisis, with total assets surpassing \$10 billion.

\smallskip
\noindent\textbf{Late 2000s to 2022.} Since 2009, SVB focused on core credit operations and experienced significant growth. Tailored financing services for large tech corporations attracted/retained major clients. Despite the challenges posed by COVID-19, SVB maintained a resilient underlying business, evident in the addition of around 1,600 new commercial clients, surpassing pre-pandemic levels. Core fee income saw substantial growth, driven by increased client investments and bolstered SVB Securities revenue. As of Q4 2022, SVB achieved great financial results, including Earnings Per Share of \$4.62, Net Income of \$275 million, and a Return on Equity of 8.9\%.

\smallskip
\noindent\textcolor{teal}{\underline{\textbf{Observation-3.\ding{202}}}: \textit{\textbf{SVB is a regional bank primarily catering to high-tech companies and has experienced rapid growth over the past 40 years. }}}

\subsection{SVB Features}\label{subsec-svb-feature}

SVB deviated from the characteristics of a typical bank (see \textcolor{teal}{Sec.\ref{subsec-tybank}}) in two notable ways.

\smallskip
\noindent\textcolor{teal}{\ding{49}} \textbf{\textit{Portfolio}}  \textcolor{teal}{[money income]} (cf. \textcolor{teal}{Fig.\ref{fig:svb_portfolio}}). SVB's portfolio exhibits a highly skewed distribution\footnote{Data source [Aug 2023]\url{https://app.dealroom.co/investors/silicon_valley_bank}}. Among its 530 customers, 62\% are technology-related firms, comprising 134 enterprise software, 80 Fintech, 39 Media, 29 Telecom, 26 Security, 11 Semiconductors, 7 Robot, and 3 Event Tech companies, showcasing its tech-centric focus. Moreover, a substantial percentage (90\%) of these companies are based in the US, indicating an unbalanced regional aspect. Additionally, SVB's venture financing predominantly centers on debt (235 Debt) and early-stage investment (69 Series A, 65 Series B, 51 Series C), constituting an outstanding 79\% and surpassing the total of other series/rounds.

\begin{figure}[!hbt]
    \centering
    \subfigure[Assets portfolio]{
    \begin{minipage}[t]{0.5\textwidth}
    \centering
    \includegraphics[width=0.9\linewidth]{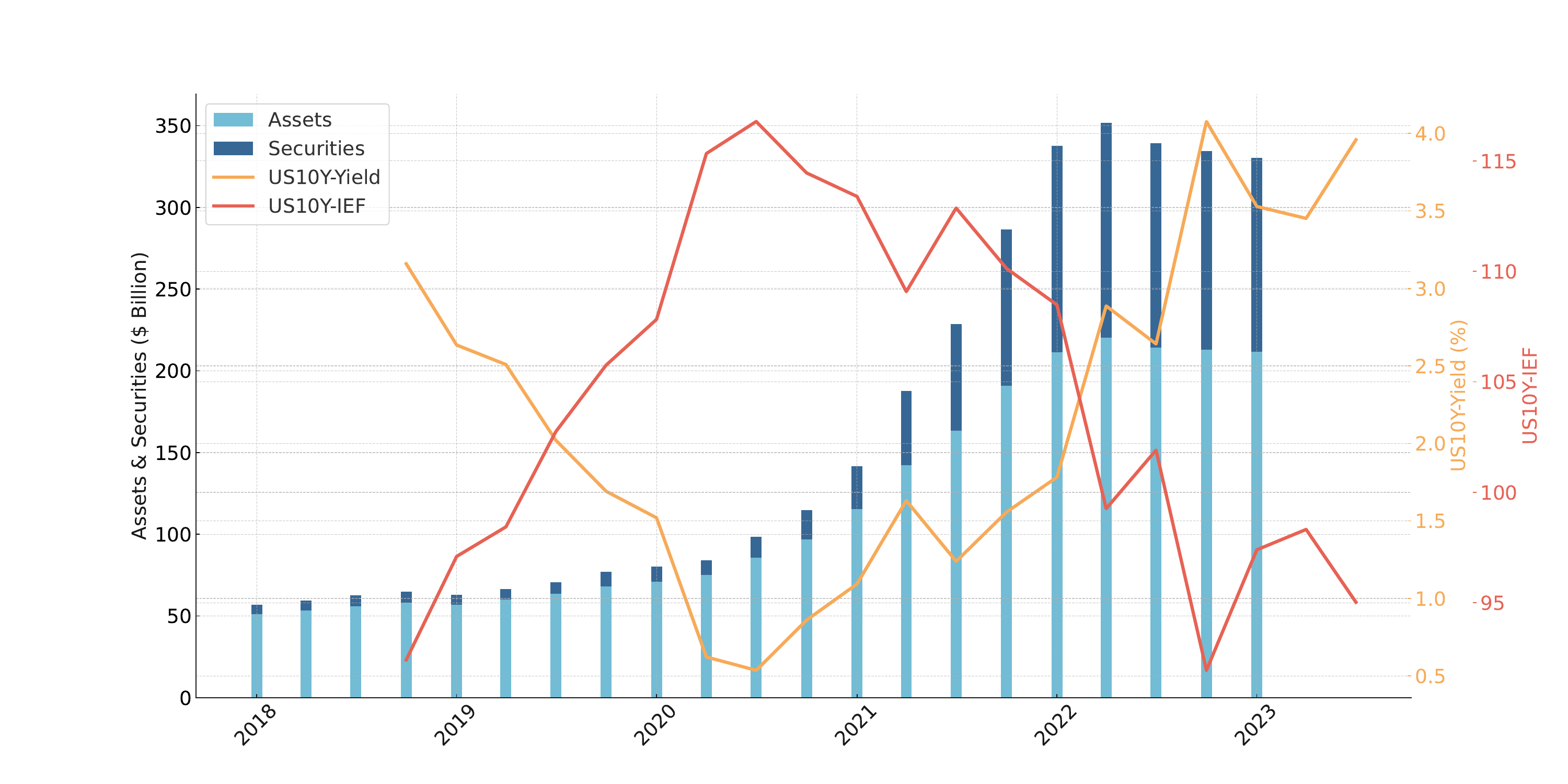}
    \end{minipage}
    \label{fig:portfolio}
    }\vspace{-0.1in}
    \subfigure[Unrealized gains (Losses)]{
    \begin{minipage}[t]{0.5\textwidth}
    \centering
    \includegraphics[width=0.9\linewidth]{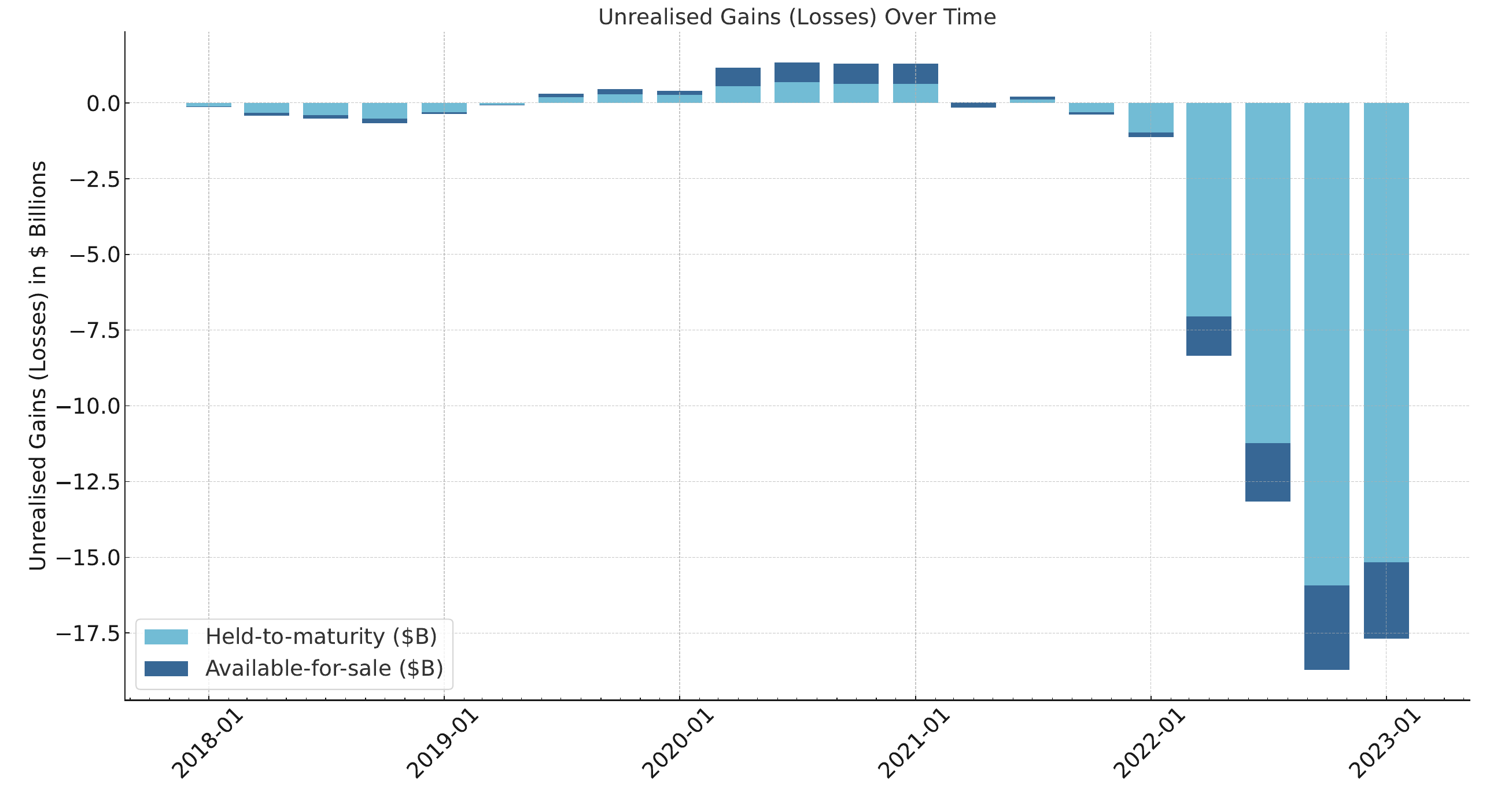}
    \end{minipage}
    \label{fig:unrealized losses}
    }\vspace{-0.15in}
\caption{Rationale of the SVB collapse}
\label{fig:svb_collapse_why}
\vspace{-0.2in}
\end{figure}

\smallskip
\noindent\textcolor{teal}{\ding{49}} \textbf{\textit{Fixed-rate government bonds}} \textcolor{teal}{[money outcome]} (\textcolor{teal}{Fig.\ref{fig:portfolio}}). SVB's assets are an undiversified concentration of its assets in the form of long-dated (10-year) fixed-rate government bonds. The downfall of SVB can be attributed to a chain of risky financial decisions. In response to the Federal Reserve's rate cuts and liquidity injections during the 2020 COVID-19 pandemic, SVB experienced a near-tripling of deposits. To capitalize on this, the bank invested heavily in long-dated fixed-rate securities and mortgage-backed securities, making it highly vulnerable to interest rate hikes. SVB further exacerbated its risk by not hedging its investments and keeping them mostly in \textit{held-to-maturity} accounts, masking the true extent of its losses.
When the Federal Reserve began aggressively raising rates in late 2021, SVB's risk exposure ballooned, culminating in losses equivalent to 150\% of its past three-decade profits. As liquidity tightened in 2022, SVB's primary clients, mostly startups, began withdrawing large sums, creating a liquidity crisis. By March 2023, forced asset sell-offs resulted in a \$1.8 billion loss, ultimately triggering a bank run and SVB's collapse.

\smallskip
\noindent\textcolor{teal}{\underline{\textbf{Observation-3.\ding{203}}}: \textit{\textbf{SVB predominantly absorbs deposits, primarily derived from its skewed portfolio, and strategically invests them in longer-term treasuries.}}}

\subsection{Factors on SVB Downfall}\label{subsec-factors}

\noindent\textcolor{teal}{\ding{49}} \textbf{\textit{Poor asset decisions}}. One of SVB's prominent assets comprised Treasury bonds with a 10-year maturity, as previously mentioned. Unfortunately, the Federal Reserve's rate decreases, particularly pronounced in 2020 as a response to the COVID-19 pandemic, had a detrimental impact on the value of these bonds, causing short-term US Treasury Yield to plummet even below those of long-term yields. SVE thus bought substantial long-term bonds. Additionally, the extended duration of long-term bonds carried heightened risks (see \textit{duration} in \textcolor{teal}{Appendix~\ref{appn-finance}}), exacerbating losses and significantly contributing to the bank's destabilization. The persistent underperformance of these Treasury bonds compounded SVB's financial challenges, further eroding confidence in its stability. In response to depositor demands, SVB was compelled to liquidate its bond holdings at a loss. The previously unrealized losses (cf. \textcolor{teal}{Fig.\ref{fig:unrealized losses}}) on the balance sheet have now become realized.

\smallskip
\noindent\textcolor{teal}{\ding{49}} \textbf{\textit{Skewed liability portfolio}}. SVB played a pivotal role as a financing source for smaller venture and private equity firms operating in the hi-tech and crypto sectors. Given a low-interest-rate environment, these startups were able to readily absorb substantial investments and subsequently deposit them with SVB. This led to a sustained growth in SVB's assets and securities (blue bars in \textcolor{teal}{Fig.\ref{fig:portfolio}}), as previously discussed, with many of these assets transitioning into long-term bonds. However, the vulnerability of these companies to financial crises increased the likelihood of them withdrawing their deposited assets whenever they sensed instability.

\smallskip
\noindent\textcolor{teal}{\ding{49}} \textbf{\textit{Weak risk control}}. Despite having rigorous loan approval procedures and broad risk prevention systems in place, SVB heavily relies on its partnerships with risk fund institutions, who provide professional insights and assist in every phase of loan management. Unfortunately, this model proved vulnerable. With the abrupt resignation of SVB's chief risk officer in 2022 and a sudden interest rate hike, SVB was unable to effectively hedge or stop loss. Further compounded by their over-reliance on Silicon Valley startups prone to mass withdrawals during crises.

\smallskip
\noindent\textcolor{teal}{\underline{\textbf{Observation-3.\ding{204}}}: \textit{\textbf{Internal factors: SVB's imbalanced portfolio, heavy dependence on long-term bonds, and inadequate risk management elevate the risk of a collapse and serve as primary contributing factors.}}}

\smallskip
\noindent\textcolor{teal}{\ding{49}} \textbf{\textit{Pressure from customers}}. The bursting of the tech bubble had a profound impact on SVB as both startup and corporate depositors faced financial challenges. This led to a significant wave of withdrawals, creating increased liability, interest rates, and liquidity risk for SVB. The rapid dissemination of news through platforms like Twitter and social media amplified the herd mentality, resulting in more jittery uninsured depositors withdrawing their funds from the bank (e.g., mass withdrawal amounting to US\$40B shortly \cite{42billion}).

\smallskip
\noindent\textcolor{teal}{\ding{49}} \textbf{\textit{Interest rate fluctuations}}. The Federal Reserve's policy shifts regarding interest rates played a central role in a sequence of events that led to SVB's collapse. Triggered by the COVID-19 pandemic, the Fed maintained a notably low-interest rate environment from 2019 to 2020, only to swiftly implement successive rate hikes that caused a substantial global financial landscape shift. This abrupt change left many bondholders without the necessary flexibility due to their lack of positioning and liquidity. Any minor disturbance had the potential to send shockwaves through the financial market, posing a significant threat to companies.

\smallskip
Therefore, as a regional bank with a distinct emphasis on high-tech startups, SVB has undergone substantial growth over the past four decades. Nevertheless, its imbalanced portfolio, with over 80\% of its customers coming from the high-tech and crypto sectors, coupled with unwise asset allocation decisions such as an influx of deposits and significant bond purchases, rendered SVB vulnerable to fluctuations in the Federal Reserve's interest rates.

\smallskip
\noindent\textcolor{teal}{\underline{\textbf{Observation-3.\ding{205}}}: \textit{\textbf{External factor: The shifts in Federal Reserve interest rates have a substantial impact on market liquidity and user confidence.}}}

%==============================================
\section{Understanding Cryptoassets}
\label{sec-cryto}
%==============================================

\subsection{Typical bank vs Cryptoassets}\label{subsec-tybank}

\noindent\textbf{Typical banks.} Banks are financial institutions that play a crucial role in facilitating the circulation of money within society. They offer a wide range of services designed to ensure the smooth functioning of financial transactions. Specifically, two major steps are,

\textit{Money source}. Banks rely on borrowing from depositors to access capital. Deposits from individuals are the main source of funds, where people deposit their money in a bank and receive interest in return. Banks also obtain funds through shareholder equity, wholesale deposits, and debt issuance. By diversifying their funding sources and focusing on captive depositors, banks increase the likelihood of retaining these funds during times of financial stress.

\textit{Make profit}. Banks determine their cost of funds based on the interest rates paid to depositors on products such as savings accounts and time deposits. They then earn money by charging higher interest rates on loans than the initial cost of funds. Mortgages, home equity loans, student loans, car loans, and credit card lending represent a significant portion of lending products, with these loans offered at variable, adjustable, or fixed interest rates.

\smallskip
\noindent\textbf{Compared to cryptoassets.}  Cryptoassets are typically based on decentralized technologies, such as blockchain, offering advanced features, including transparency, immutability, and decentralization. Differences between them lie in the nature of their design goals.

\textit{Intermediaries}. Banks act as intermediaries (equiv. trusted third parties) in financial sectors to ensure the security and integrity of transactions. In contrast, cryptoassets enable peer-to-peer transactions directly between individuals without intermediaries. Transactions are validated and permanently recorded on-chain.

\textit{Regulation}. Banks are subject to strict oversight by financial authorities. They are required to comply with Know Your Customer (KYC) and Anti-Money Laundering (AML) regulations to prevent illicit activities. Cryptoassets, being relatively decentralized,  face varying degrees of regulatory scrutiny in different jurisdictions~\cite{wahrstatter2023blockchain}.

 \textit{Fiat currency}. Traditional banks primarily deal with fiat currency, which is government-issued money such as the US dollar or Euro. They hold customer deposits in fiat currency and facilitate transactions denominated in these currencies. Cryptoassets, on the other hand, are digital assets that exist solely in electronic form. They are not backed by any physical asset (stablecoins \cite{moin2020sok} excluded), and their value is determined by market demand and supply dynamics.

\smallskip
\noindent\textcolor{teal}{\underline{\textbf{Observation-4.\ding{202}}}: \textit{\textbf{Cryptoassets distinguish from traditional government-issued currencies through their lack of regulation, absence of reserves, and autonomous insurance. These cryptoassets function as universal equivalents within the crypto space, giving rise to a multitude of digital products.}}}

\subsection{Cryptoasset Classification} \label{subsec-classify}

\smallskip
\noindent\textbf{Native tokens.} Native tokens are the inherent cryptocurrencies of specific blockchain networks, serving as the primary medium of transaction within their respective environments. They are frequently used to incentivize network participation, pay transaction fees,  enable unique network functions and maintain the normal operation of chains. Examples include BTC on the Bitcoin network, ETH on Ethereum and BNB on the Binance smart chain (BSC). A series of investment companies (e.g., JP Morgan, Goldman Sachs, A16z) have adopted BTC or ETH as alternative reserves to maintain liquidity and mitigate similar risks in traditional banking systems.

\smallskip
\noindent\textbf{X-20 tokens.} X-20 tokens are standard-compliant tokens issued on blockchain platforms that support functions ranging from simple scripting to complicated smart contracts, such as Bitcoin and Ethereum that are supported by BRC-20~\cite{binancebrc} and ERC-20~\cite{erc20}, respectively. These tokens can represent a wide array of digital or physical assets and are often used in decentralized applications (DApps) \cite{antonopoulos2018mastering} and decentralized finance (DeFi) protocols \cite{werner2022sok}.

\smallskip
\noindent\textbf{Crypto stablecoin.}
Stablecoins are cryptocurrencies designed to preserve stable value by pegging to specific assets \cite{moin2020sok,fu2023rational}. Stablecoins are commonly pegged to traditional fiat currencies such as US dollars or other assets such as gold. For example, Tether (USDT) \cite{lipton2020tether} and USD Coin (USDC) \cite{fiedler2023stablecoins} are stablecoins pegged to the US dollar, providing a balance of cryptocurrency benefits such as transaction speed and privacy with the stable value of fiat currencies.

\smallskip
\noindent\textbf{Derivatives.} 
Crypto derivatives are financial contracts that derive their value from underlying cryptocurrency assets. They provide investors with the opportunity to speculate on the future price movements of the underlying asset without owning it directly. There are two major types of crypto derivatives. The first type operates within traditional financial markets, where cryptocurrencies serve as underlying assets for creating financial products. Examples of such derivatives include Bitcoin futures\textcolor{gray}{\small\&}options, which are traded on platforms like the Chicago Mercantile Exchange (CME)~\cite{cmegroup}. These derivatives function as both risk management tools and speculative instruments within the cryptoasset market. The second type involves building financial products directly on blockchain platforms, also known as DeFi \cite{werner2022sok}. Instead of relying on centralized agencies, financial products are developed using smart contracts on blockchain networks. In-use examples are detailed in Appendix~\ref{appn-defi}

\smallskip
\noindent\textcolor{teal}{\underline{\textbf{Observation-4.\ding{203}}}: \textit{\textbf{As a user ventures into the cryptocurrency space, they typically encounter four distinct categories of cryptoassets progressively: stablecoins, serving as a gateway for exchanging real-world assets (RWA) into crypto tokens, which includes native platform tokens and X-20 tokens. Platform tokens can be utilized to create additional X-20 tokens, which can, in turn, be used to create a variety of derivatives that enable further composability and financial innovation.}}}

\subsection{Technical Supports}\label{subsec-decouple}

\noindent\textbf{Blockchain/Ledger.} Blockchain functions as a decentralized digital ledger that ensures secure and transparent recording of transactions \cite{bonneau2015sok}. Transactions are grouped into blocks and arranged in chronological order, forming a chain-like structure known as the blockchain. Adding a new block to the chain requires consensus among a wide range of participants. Consensus procedures, such as Proof of Work (PoW) \cite{garay2020sok}, rely on reliable randomness generation to determine the next block proposer. Participants must agree on the rules for becoming proposers and the associated outcomes. This process is driven by economic incentives that provide miners or validators with rewarding incentives for their participation.

\smallskip
\noindent\textbf{Smart contract.} A smart contract represents a unique type of contract in which the terms of the agreement are converted directly into executable code. This computational entity functions as a self-sustaining state machine \cite{li2022smart}, ensuring a harmonious interaction of inputs and outputs and circumventing the need for reliable third-party intermediaries. The contracts offer transparency, security, and efficiency as well. Primarily facilitated on blockchain infrastructures such as Ethereum, smart contracts undergo automatic execution upon the satisfaction of pre-established conditions. The adaptability of smart contracts fosters automation across a wide array of sectors, ranging from financial services~\cite{werner2022sok}, decentralized administrative~\cite{yu2022leveraging,wang2022empirical} and a series of applications \cite{antonopoulos2018mastering}. 

\smallskip
\noindent\textbf{Token standard.}   
Tokens are subject to specific standards that outline the procedures for their creation, deployment, and issuance. These standards ensure interoperability and compatibility among different tokens and facilitate seamless integration within blockchain systems. Ethereum has risen to prominence in the realm of token standards and facilitating a diverse range of applications within its ecosystem. The fungible ERC-20~\cite{erc20} and non-fungible ERC-721~\cite{wang2021non} token standards are salient examples of Ethereum's influence, having catalyzed the expansion of the token marketplace \cite{wang2022exploring}. Similar token standards serve as foundational templates and expand within the Binance~\cite{bep20} and Avalanche~\cite{arc20} ecosystems.

\smallskip
\noindent\textbf{Economic incentives.}
Economic incentives play a pivotal role in motivating individuals to participate in blockchain networks. Users have various avenues to earn rewards, with the primary methods being through \textit{formulaic rewards} and \textit{market arbitrage}. Formulaic rewards are tied to consensus mechanisms such as PoW \cite{nakamoto2008bitcoin} and PoS \cite{king2012ppcoin}. In PoW, nodes (miners) are incentivized to solve complex mathematical problems, verify transactions, and add new blocks to the blockchain. In PoS, nodes (validators) are incentivized based on their stake in the system, receiving economic incentives each time they successfully validate and add a block. Market arbitrage offers another avenue for profit, involving expertise in DeFi protocols and capitalizing on price discrepancies. This approach is akin to arbitrageurs in traditional financial markets. For blockchain developers or early employees, participating in a project team allows them to benefit from the texit{initial funds raised} and receive tangible rewards in the form of tokens or other forms of compensation.

\smallskip
\noindent\textcolor{teal}{\underline{\textbf{Observation-4.\ding{204}}}: \textit{\textbf{Cryptoassets operate within blockchain ecosystems, heavily reliant on the resilience of these platforms. X-20 tokens and stablecoins, in particular, hinge upon smart contract-supported blockchains and adhere to their token standards. Users are drawn to crypto space due to decentralization, borderless trading, and the potential for lucrative incentives.}}}

\begin{figure*}[!htb]
    \centering
    \subfigure[BTC]{
    \begin{minipage}[t]{0.32\textwidth}
    \centering
    \includegraphics[width=2.6in]{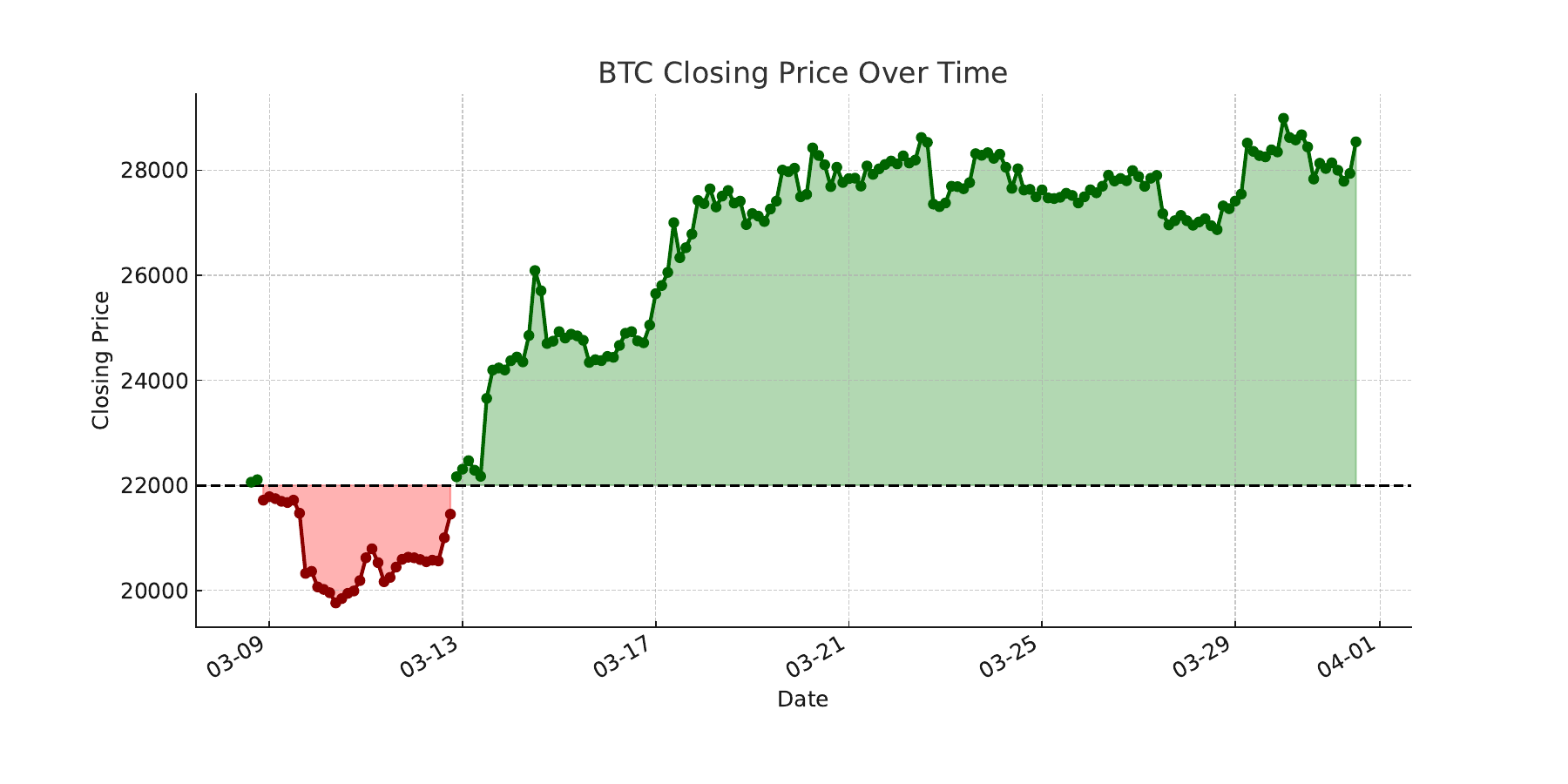}
    \end{minipage}
    \label{fig:closing_price_btc}
    }
    \subfigure[LTC]{
    \begin{minipage}[t]{0.32\textwidth}
    \centering
    \includegraphics[width=2.6in]{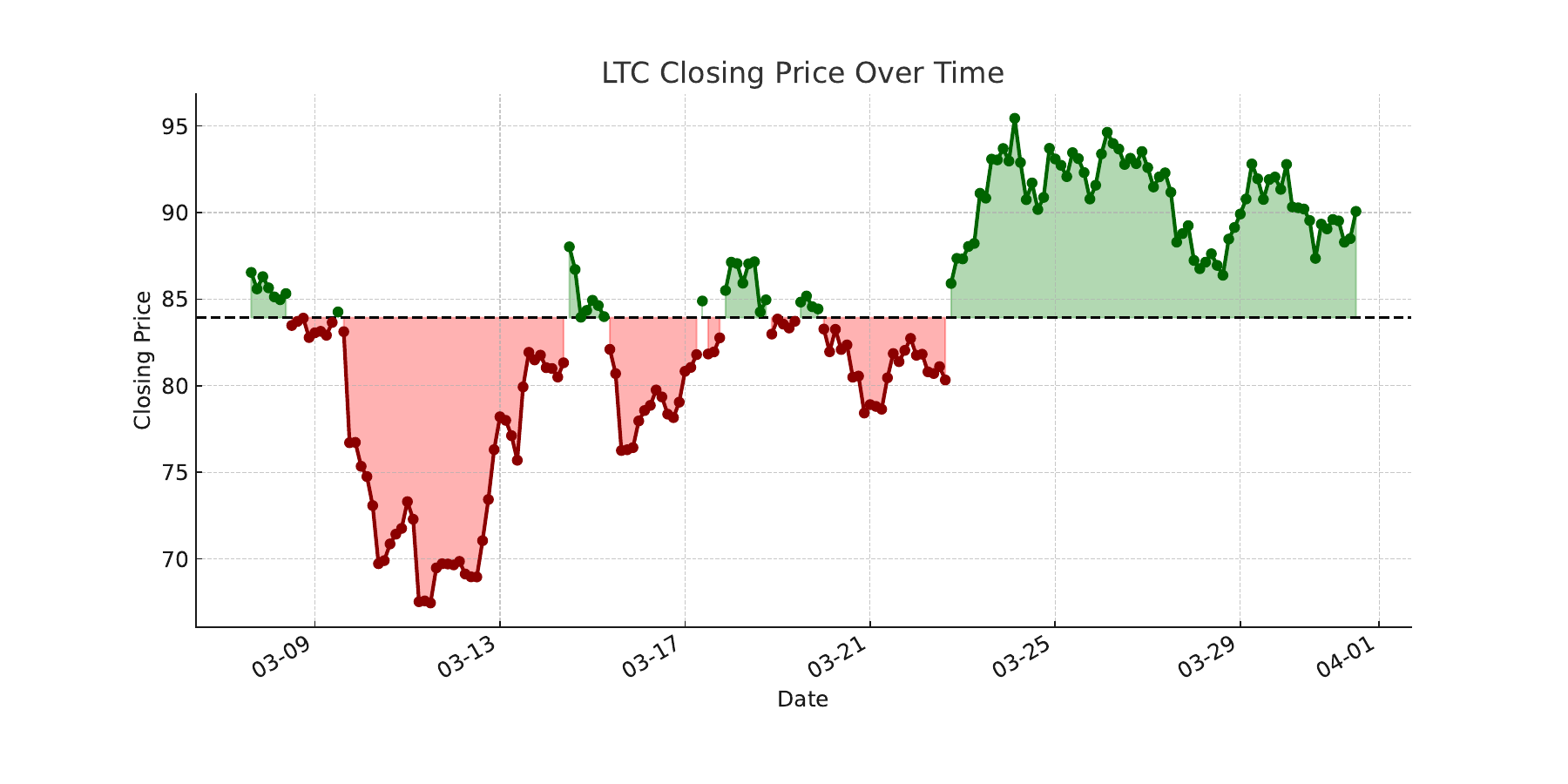}
    \end{minipage}
    \label{fig:closing_price_ltc}
    }
    \subfigure[DOGE]{
    \begin{minipage}[t]{0.32\textwidth}
    \centering
    \includegraphics[width=2.6in]{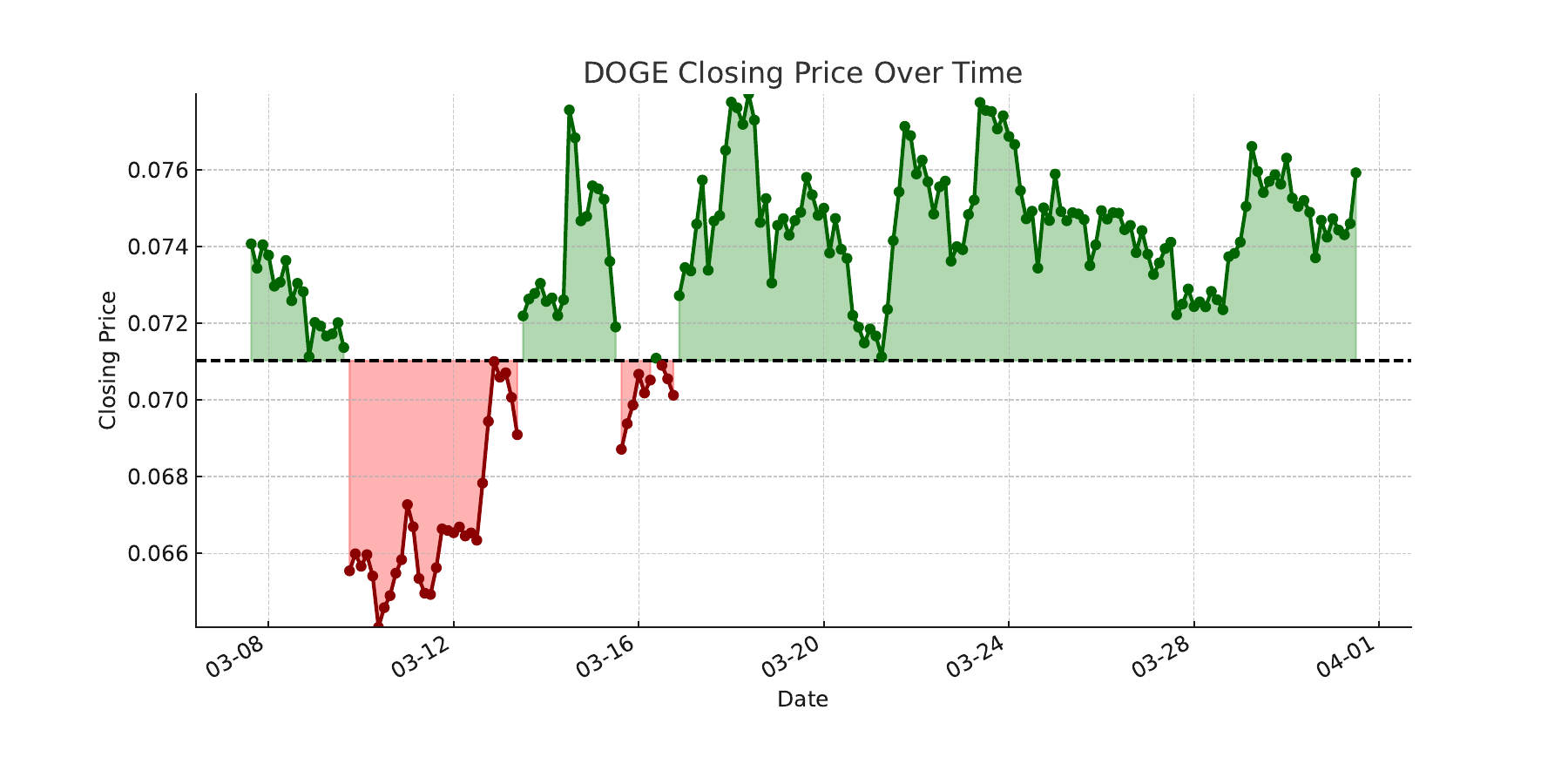}
    \end{minipage}
    \label{fig:closing_price_doge}
    }\vspace{-0.15in}
    \subfigure[ETH]{
    \begin{minipage}[t]{0.32\textwidth}
    \centering
    \includegraphics[width=2.6in]{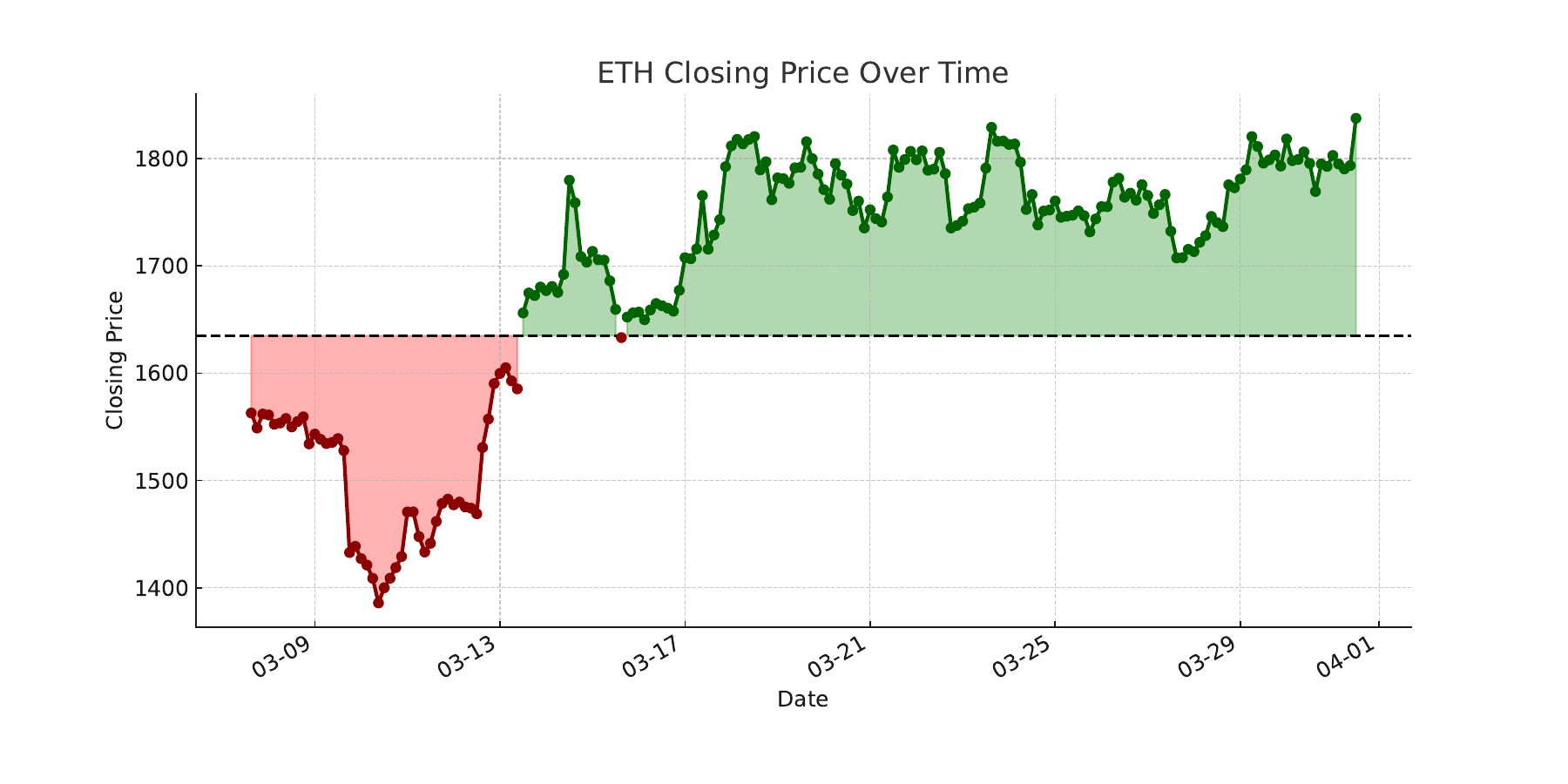}
    \end{minipage}
    \label{fig:closing_price_eth}
    }
    \subfigure[AVA]{
    \begin{minipage}[t]{0.32\textwidth}
    \centering
    \includegraphics[width=2.6in]{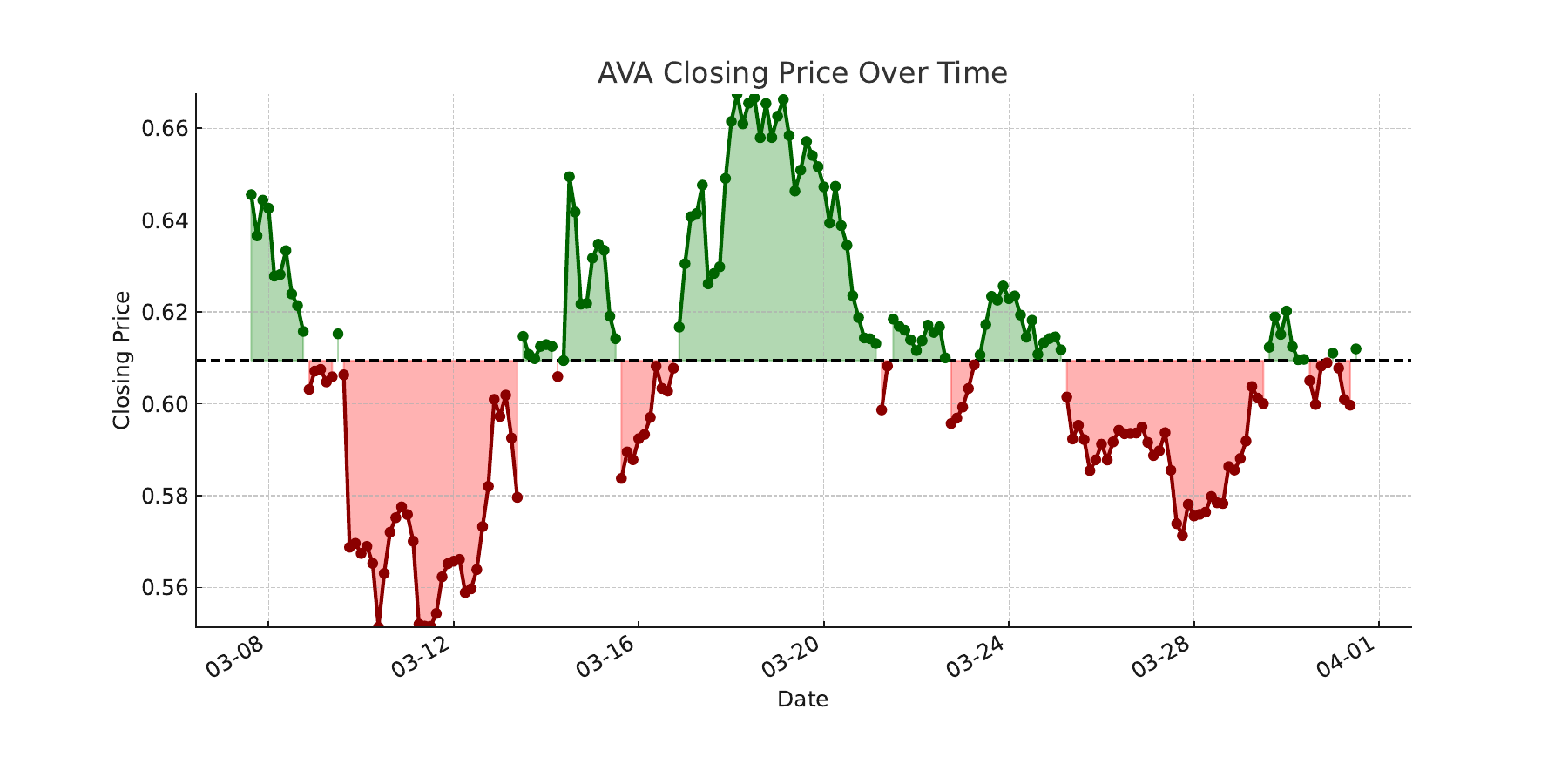}
    \end{minipage}
    \label{fig:closing_price_ava}
    }
    \subfigure[BNB]{
    \begin{minipage}[t]{0.32\textwidth}
    \centering
    \includegraphics[width=2.6in]{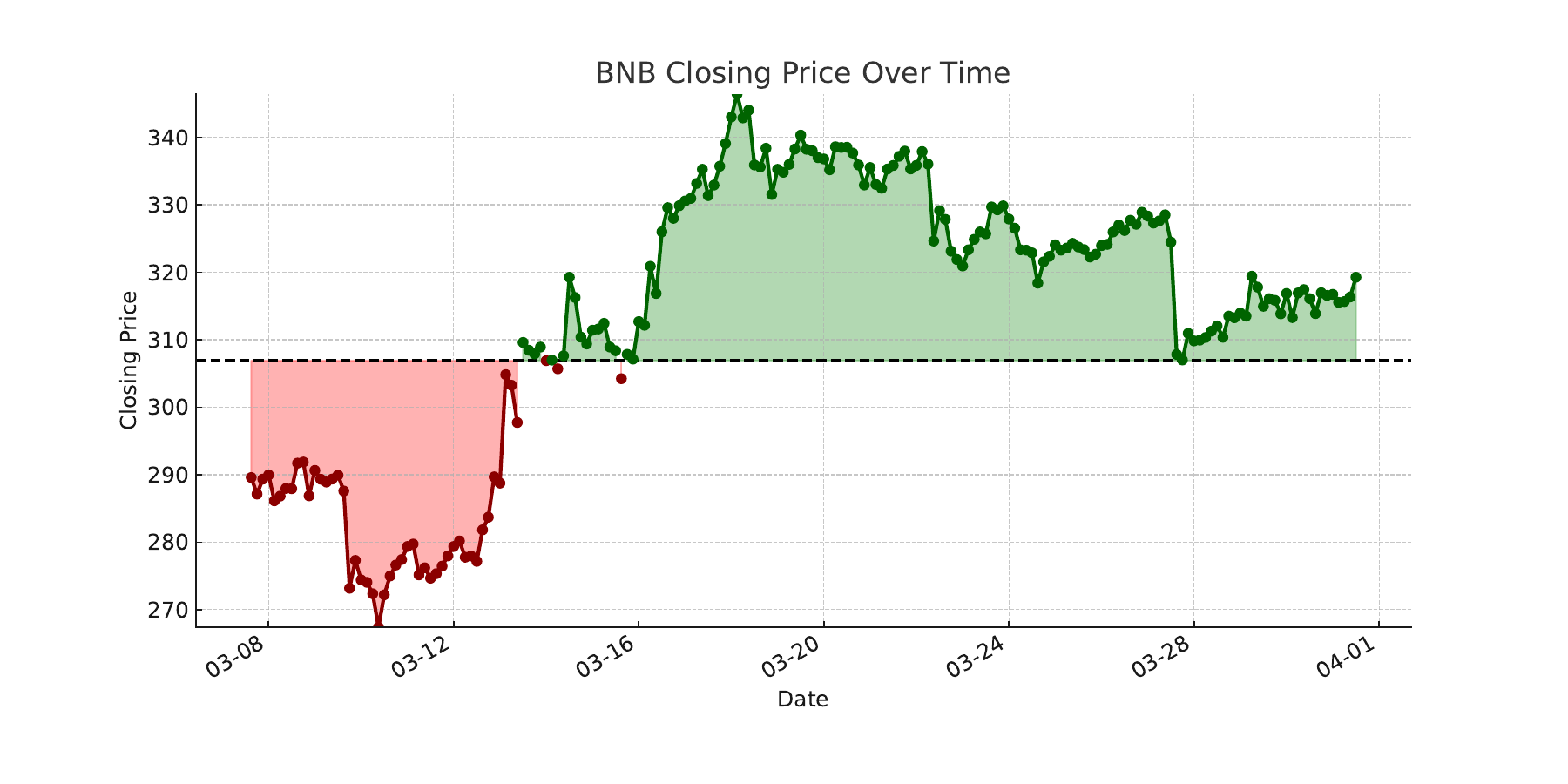}
    \end{minipage}
    \label{fig:closing_price_bnb}
    }\vspace{-0.15in}
    \subfigure[USDC]{
    \begin{minipage}[t]{0.32\textwidth}
    \centering
    \includegraphics[width=2.6in]{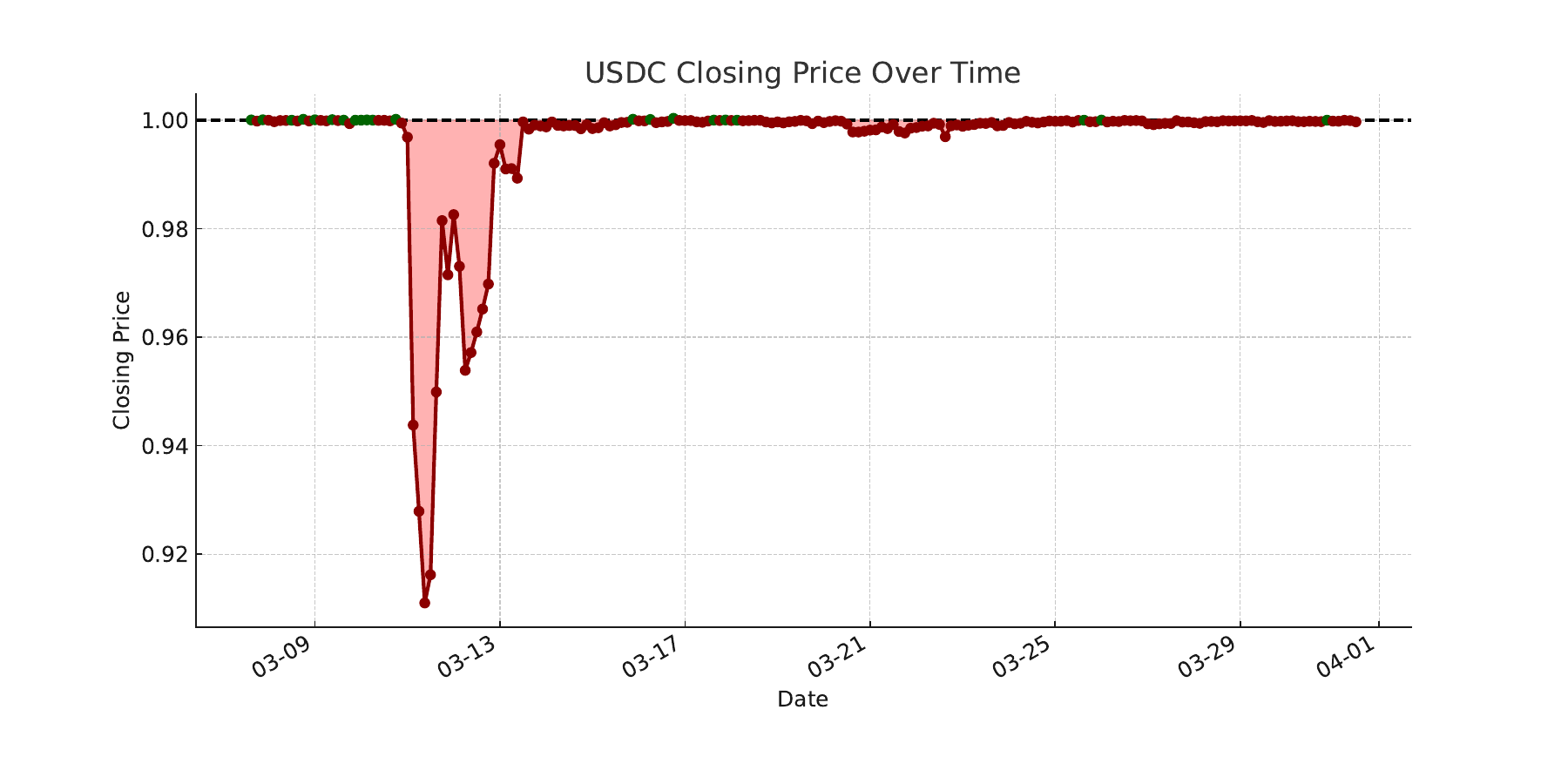}
    \end{minipage}
    \label{fig:closing_price_usdc}
    }
    \subfigure[USDT]{
    \begin{minipage}[t]{0.32\textwidth}
    \centering
    \includegraphics[width=2.6in]{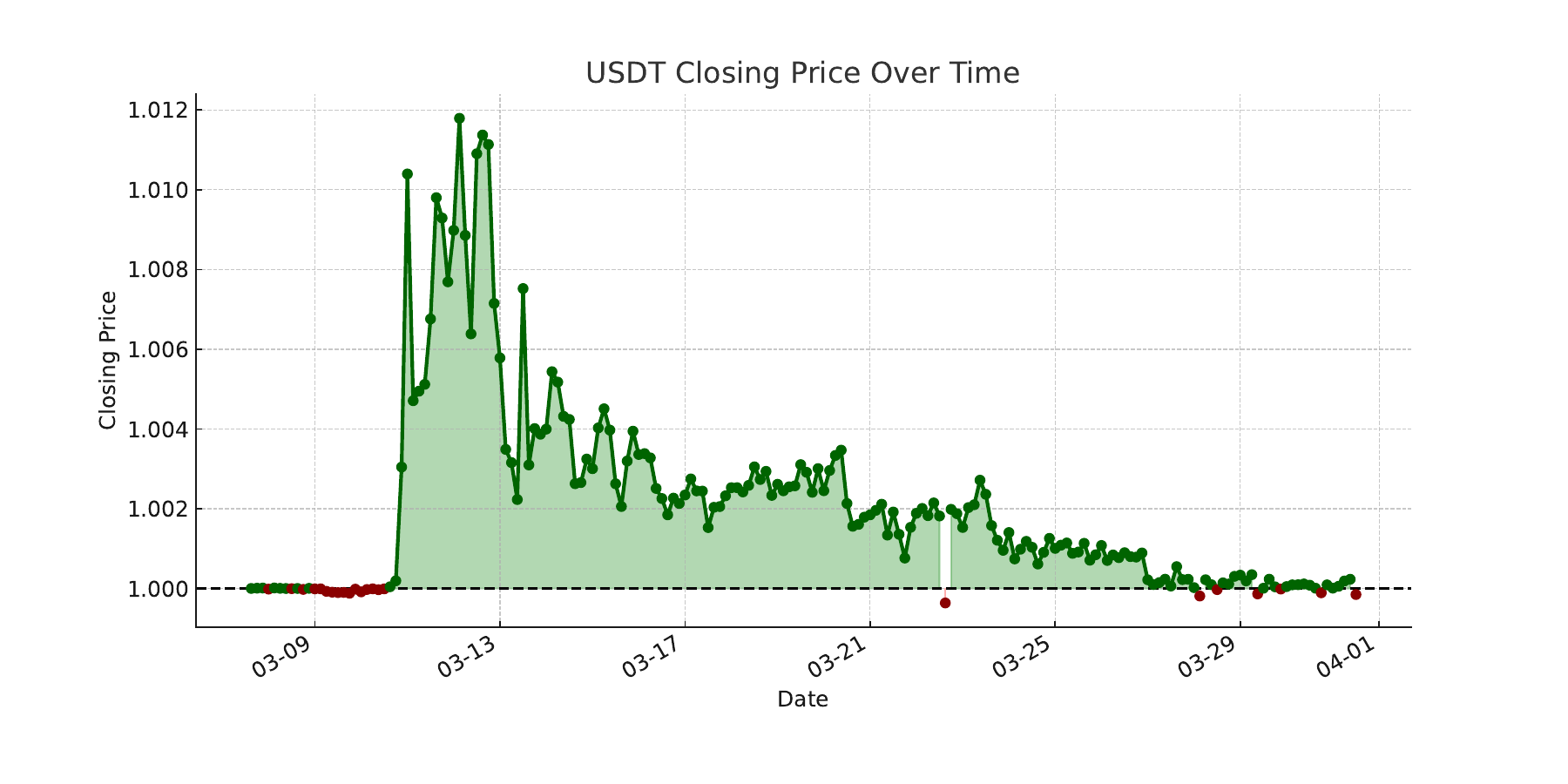}
    \end{minipage}
    \label{fig:closing_price_usdt}
    }
    \subfigure[BUSD]{
    \begin{minipage}[t]{0.32\textwidth}
    \centering
    \includegraphics[width=2.5in]{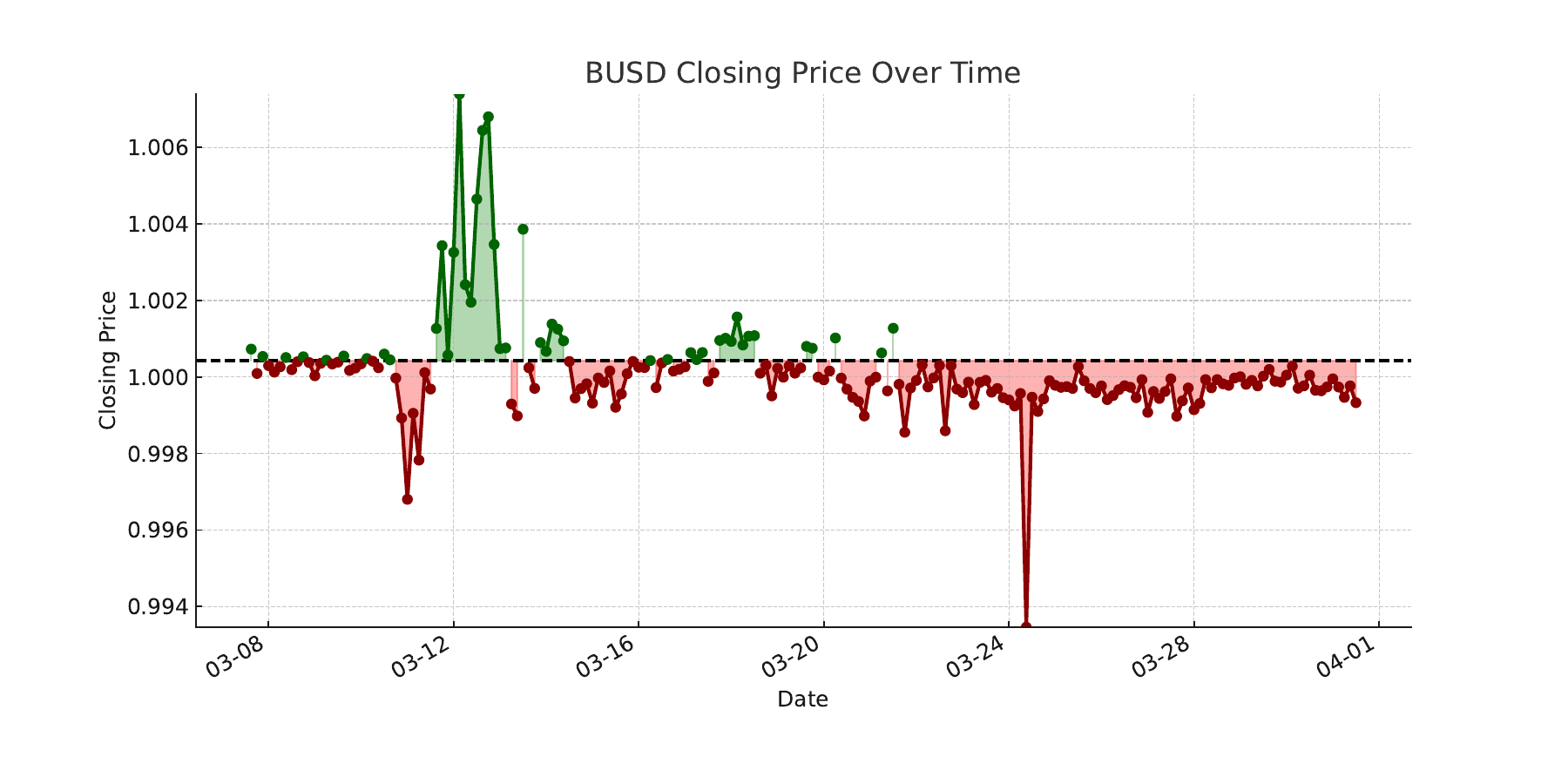}
    \end{minipage}
    \label{fig:closing_price_busd}
    }
    \vspace{-0.15in}
\caption{The price trend of prevalent cryptocurrencies (March 09 - April 01, 2023).}
\label{fig:closing_price}
    \vspace{-0.15in}
\end{figure*}

%=================================================  
\section{Empirical Study}
\label{sec-emprical}
%=================================================  

\subsection{Cryptoassets Price Trends}\label{subsec-crypto-price}

\noindent\textcolor{teal}{\textit{\textbf{Glance.}}} We carefully curate a selection of representative projects, including both native blockchain tokens and stablecoins. The native tokens contain two types: Bitcoin-like tokens (BTC, LTC, DOGE) represent projects launched around the same time as Bitcoin, with their chain code largely mirrored from Bitcoin's GitHub repository. Ethereum-like tokens (ETH, BNB, AVA) refer to platforms that are built on Ethereum's design and implementation, offering full functionalities such as smart contracts. Stablecoins represent an orthogonal dimension of crypto-tokens that are independent of crypto-tokens but are pegged to fiat currencies (e.g., USD). Importantly, X-20 tokens are generated without any underlying assets, rendering them of lesser value and possessing limited functionality. As such, we have excluded them from our analysis. It is noteworthy that all the referenced tokens we have included in our study are among the top 10 in terms of market capitalization. Our data covers a one-month period, starting from early March (before the SVB collapse) and concluding in early April (following SVB's acquisition).

\smallskip
\noindent\textbf{Bitcoin-like tokens} (\textcolor{teal}{Fig.\ref{fig:closing_price_btc}}-\textcolor{teal}{\ref{fig:closing_price_doge}}).
The price of BTC exhibited fluctuations during this period. On March 7th, it stood at \$22,338.81, only to experience a drop to around \$19,500, reflecting the ripple effect influenced by SVB's market events. Similarly, LTC's price also fluctuated during March. DOGE's price trend during this period was influenced by the same macroeconomic factors, along with its meme-driven nature, resulting in more pronounced fluctuations than other cryptocurrencies.

An intriguing observation is that the prices of these tokens recovered to the same level of prices after three days (Mar 9 - 12). Subsequently, prices saw a slight increase and exhibited signs of recovery. This reflects the SVB collapse and its initial impact on overall market confidence. While the collapse initially shook confidence, the subsequent impact proved to be limited. The prices of mainstream crypto-tokens demonstrated resilience in the face of this regional bank collapse.

\smallskip
\noindent\textbf{Ethereum like tokens} (\textcolor{teal}{Fig.\ref{fig:closing_price_eth}}-\textcolor{teal}{\ref{fig:closing_price_ava}}).
The price behavior of ETH closely resembles that of BTC, influenced by the overall trend in crypto markets but demonstrating a quick recovery. In contrast, the AVA price trend reflects a broader market dynamic. It initially increased back to its price but then experienced a subsequent decline. The SVB collapse is believed to have contributed to this uncertainty. BNB, being a part of a major crypto exchange, might have been expected to be significantly impacted by the overall liquidity crunch in the banking system. Interestingly, the data indicate its rapid recovery from this event. This suggests that conventional banks have limited impact on crypto exchanges, primarily due to differences in their operational strategies and the composition of customer bases.

\smallskip
\noindent\textcolor{teal}{\underline{\textbf{Observation-5.\ding{202}}}: \textit{\textbf{Major cryptoassets display a strong correlation, mirroring each other's price movements. While undergoing steep declines during market downturns, they tend to rebound swiftly, reclaiming their pre-collapse price levels.}}}

\smallskip
\noindent\textbf{Stablecoins} (\textcolor{teal}{Fig.\ref{fig:closing_price_usdc}}-\textcolor{teal}{\ref{fig:closing_price_busd}}). As stablecoins, USTC, USDT, and BUSD are expected to maintain a close peg to \$1. However, the SVB collapse and its impact on the USDC stablecoin have raised concerns about the overall stability of stablecoins. During March 9-12, all three of these stablecoins experienced a sharp deviation, de-pegging from their anchored tokens by as much as 10\% (typically staying within a 1\% range).
Subsequently, they all demonstrated a recovery. The difference lies in the pace of stabilization. USDC quickly returned to its regular threshold, while USDT took around 20 days to stabilize within a narrow range. BUSD, on the other hand, experienced a more prolonged fluctuation, albeit within a relatively narrow 3\% range.

\smallskip
\noindent\textcolor{teal}{\underline{\textbf{Observation-5.\ding{203}}}: \textit{\textbf{Stablecoins confront the risk of deviating from the peg, but return to stable fluctuation range quickly.}}}

%===================================
\subsection{Sentiment in Tweets}\label{subsec-sentiment-tweet}

\begin{figure}[!htb]
    \centering
    \subfigure[The number of \textit{\#SVB}-related tweets over time]{\label{fig:tweets_svb}
    \begin{minipage}[t]{1\linewidth}
    \centering
    \includegraphics[width=1\linewidth]{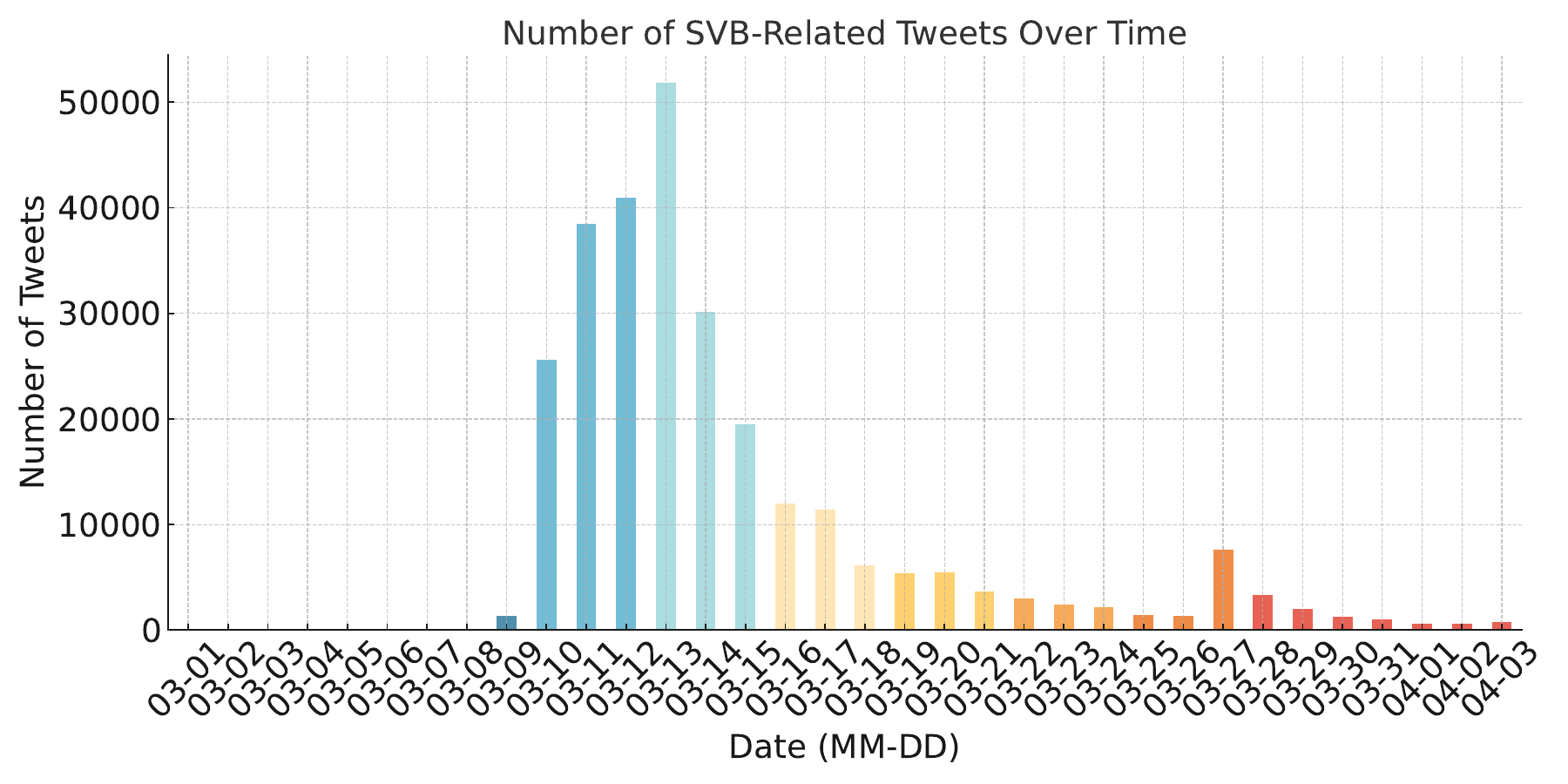}
    \end{minipage}
    }\vspace{-0.15in}
    \subfigure[The number of \textit{\#Crypto} tweets over time]{\label{fig:tweets_crypto}
    \begin{minipage}[t]{\linewidth}
    \centering
    \includegraphics[width=1.1\linewidth]{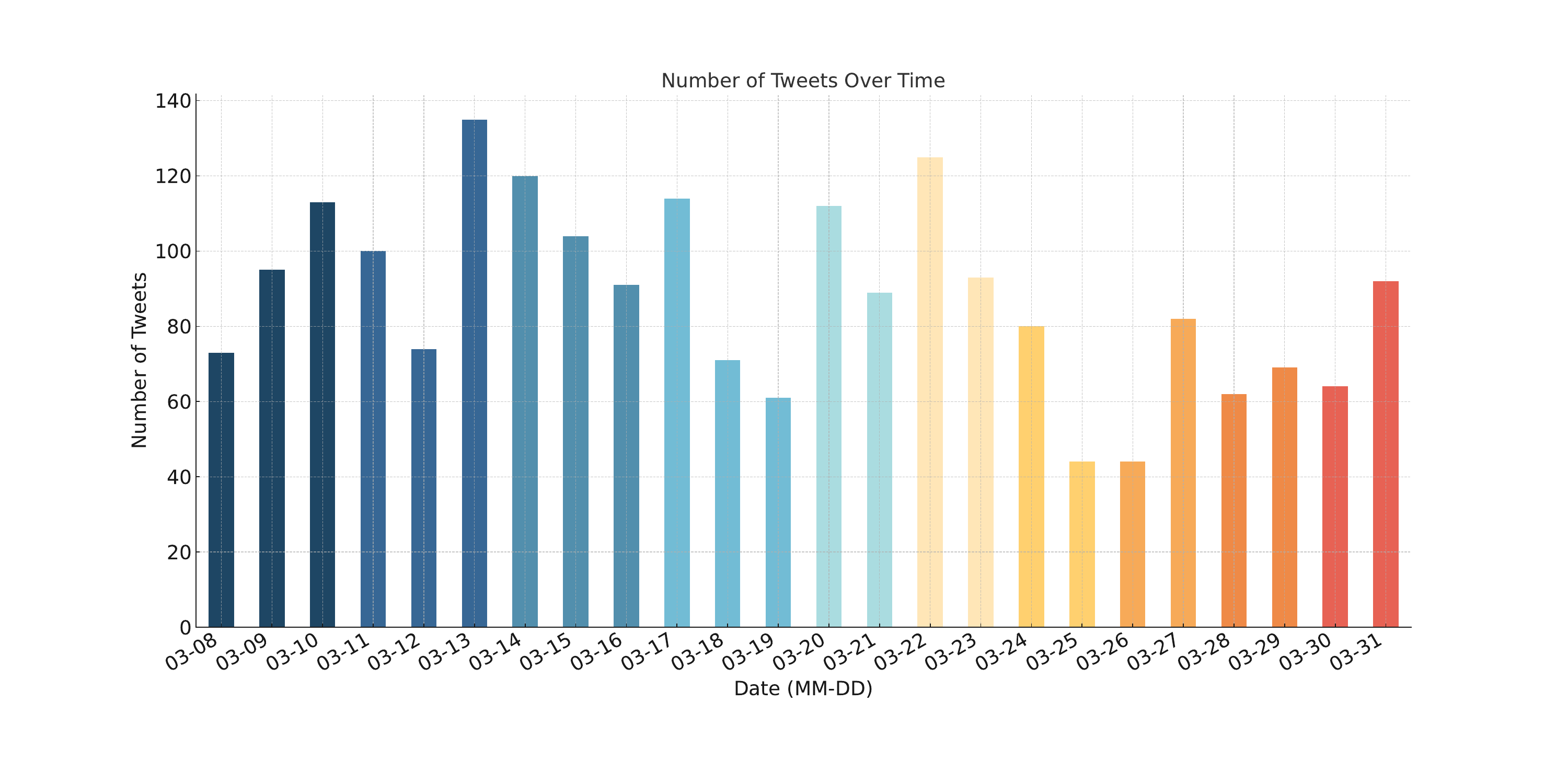}
    \end{minipage}
    }\vspace{-0.1in}
    \subfigure[Sentiment Analysis of \textit{\#Crypto} on Twitters]{\label{fig:sentiment_crypto}
    \begin{minipage}[t]{\linewidth}
    \centering
    \includegraphics[width=1\linewidth]{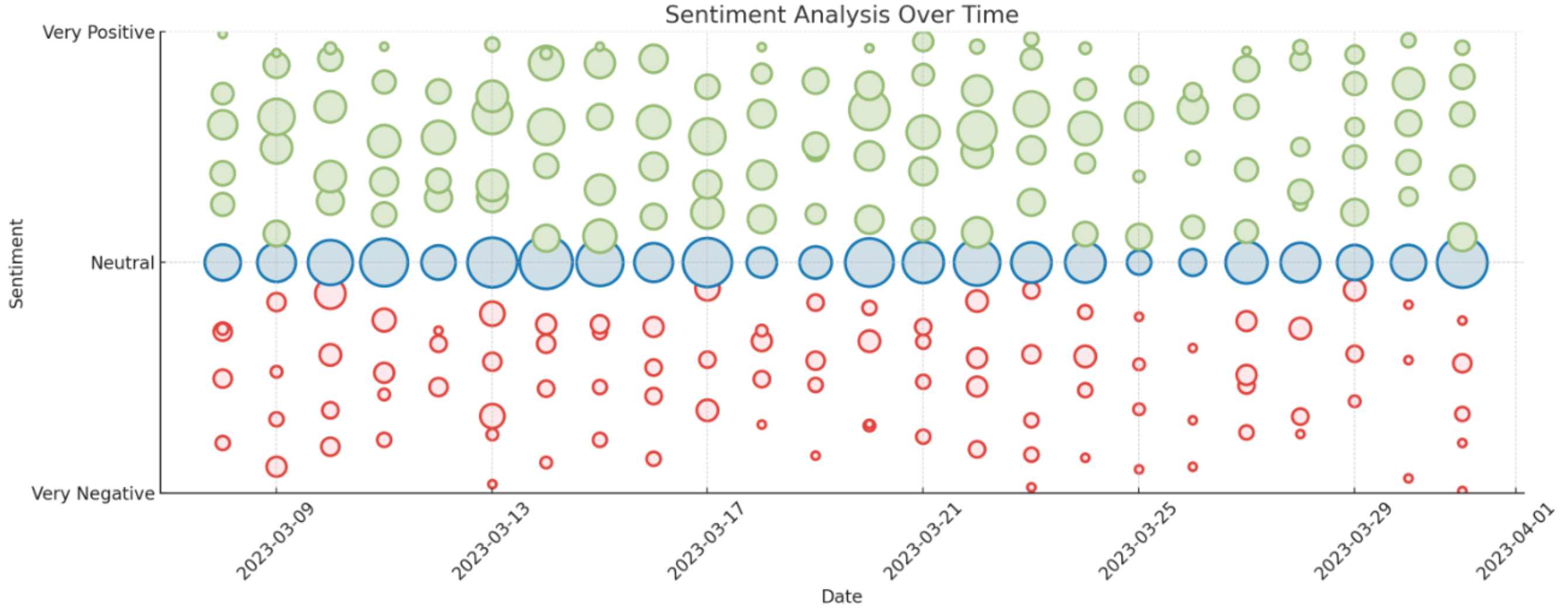}
    \end{minipage}
    } \vspace{-0.15in}
    \caption{Sentiment analyses by tweets over time}
    \label{fig:tweets}
    
\end{figure}

\noindent\textcolor{teal}{\textit{\textbf{Glance.}}} We have conducted a series of analyses of social sentiment surrounding SVB's collapse and the discussions related to cryptocurrencies during the same timeframe (\textcolor{teal}{Fig.\ref{fig:tweets}}). We gathered publicly available data from the widely used social platform X (formerly Twitter). To streamline our analysis, we focused on hashtags with the most representative keywords, including \textit{\#SVB collapse} (as well as related terms like SVB failure, falldown, etc.) and \textit{\#Crypto}. The data covers a one-month period, spanning the entirety of March, which coincides with the collapse of SVB. Based on the received crypto-related tweets, we further conducted a content analysis

\smallskip
\noindent\textbf{Sentiment in SVB} (\textcolor{teal}{Fig.\ref{fig:tweets_svb}}). Throughout March, a significant increase in discussions (tweets) related to SVB was observed. Prior to its collapse date, there were negligible tweets on this topic, but after the collapse, the number of tweets surged to over 50,000 at its peak. However, user attention quickly diminished afterward, suggesting that the impact was limited and short-term. This observation is consistent with the expectation that in the event of a widespread financial crisis resulting from the SVB collapse, public attention would typically continue for several months.

\smallskip
\noindent\textbf{Sentiment in Crypto} (\textcolor{teal}{Fig.\ref{fig:tweets_crypto}}).  
Comparatively, the volume of tweets related to SVB significantly outweighs the discussions around \textit{\#Crypto}, which maintain a consistently low level, averaging around 100+. However, when considering its sub-keywords like \textit{\#ETH}, \textit{\#BTC}, etc., this figure is expected to rise to 700+. It's noteworthy that tweet users appear to engage in relatively fewer discussions regarding their crypto assets, or at least, they rarely mention both SVB and their crypto holdings in the same tweet. This suggests an indeterminate causal relationship, which defies our intuitive expectation that such a significant collapse may merely generate limited substantial concerns among crypto users.

\smallskip
\noindent\textcolor{teal}{\underline{\textbf{Observation-5.\ding{204}}}: \textit{\textbf{The volume of crypto-related tweets does not seem to explicitly correlate with an increase in tweets related to the SVB collapse, even during periods of heightened SVB collapse-related tweet activity.}}}

\smallskip
\noindent\textbf{Content analysis} (\textcolor{teal}{Fig.\ref{fig:sentiment_crypto}}). Following our received tweets, we conducted the content analysis and classified sentiments into three distinct categories: positive, neutral, and negative. The outcomes of this analysis highlight a wide-ranging distribution of sentiments.

The green circles, representing \textit{positive} sentiments, exhibit some concentration during specific periods. Larger green circles indicate a higher frequency of similar sentiment levels. Tweets in this category often express optimism, with users stating phrases like ``holding cryptocurrencies!'' or ``crypto will survive long'' or ``recover soon''. These sentiments reflect positive news and an optimistic outlook circulating within the market.

The blue circles, denoting \textit{neutral} sentiments, are predominantly centered around the zero mark. Larger blue circles might signify collective ambivalence or users awaiting clear signals from the market. The presence of neutral sentiments indicates periods of uncertainty or a lack of strong opinions. Many tweets in this category express sentiments such as ``wish everything will be okay'', ``will CEX crash?'' or ``should I swap my crypto?''

The red circles, representing \textit{negative} sentiments, also exhibit clustering, although their absolute size is smaller than the other two types. A few large red circles might align with significant events or news that triggered dissatisfaction or concern. Users in this category expressed sentiments such as ``escape!'' or ``have cleared out all my stakes''. The SVB collapse and its impact on the crypto markets contribute to negative sentiments, especially concerning unstable stablecoins and the broader financial ecosystem.

\smallskip
\noindent\textcolor{teal}{\underline{\textbf{Observation-5.\ding{205}}}: \textit{\textbf{Twitter users generally exhibit more positive behavior than negative regarding the collapse.}}}

%===================================
\subsection{Sentiment Analysis}\label{subsec-senti-composite}

\begin{figure}[!htb]
    \centering
    \includegraphics[width=0.9\linewidth]{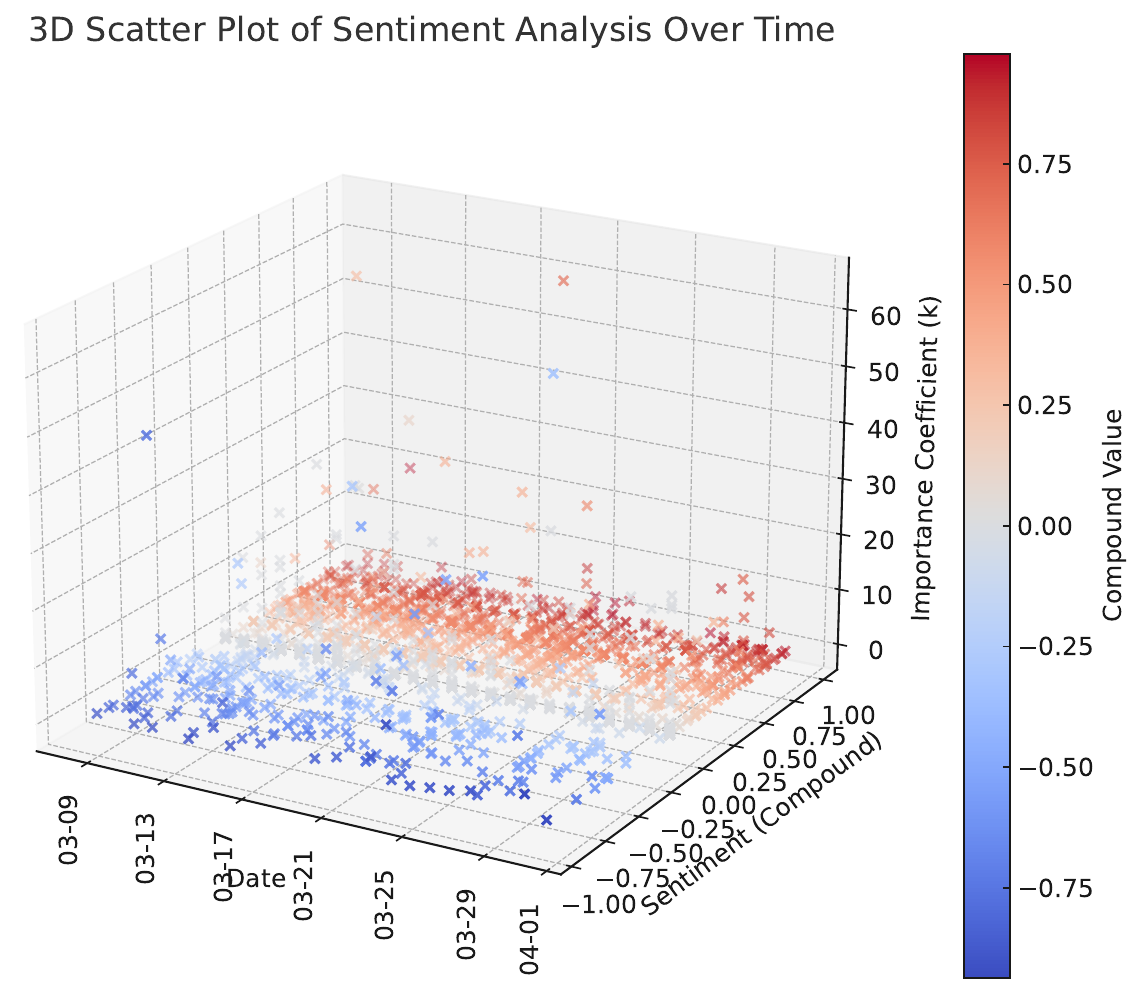}
    \caption{Sentiment analysis of \textit{\#Crypto} (Importance Coefficient)}
    \label{fig:sentiment_3d}
\end{figure}

\noindent\textcolor{teal}{\textit{\textbf{Glance.}}} We conducted a more in-depth sentiment analysis, considering multiple dimensions. Building upon the previous data, we examined the relationship between the date, importance coefficient, and sentiment (compound score) of tweets related to cryptocurrencies (\textcolor{teal}{Fig.\ref{fig:sentiment_3d}}). In this analysis, the dimension known as the importance coefficient indicates the influence level of authors and reveals the role of influencers in shaping sentiment. For clarity, we have divided the timeframe into three distinct stages for analysis: pre-collapse, during the collapse, and post-collapse. Each stage spans approximately ten days, resulting in a total duration of one month (March).

\smallskip
\noindent\textbf{Pre-Collapse.} Leading up to the SVB collapse, sentiment remained relatively stable with minor fluctuations. It is inferred that some influencers may have sensed or received key information from alternative sources, prompting early signals. The importance coefficient here reflected author influence and indicated a mixture of sentiments among both influential and less influential authors.

\smallskip
\noindent\textbf{SVB Collapse.} A significant shift in sentiment occurred in response to the SVB collapse event. Negative sentiments became more prominent, particularly among influential authors (as indicated by higher importance coefficients).  This shift directly mirrors the impact of SVB's failure and the subsequent ripple effect on the crypto industry. The strategic move of funds by entities like Circle, and public opinions from well-known figures in the industry, also contribute to the sentiment trend. Concerns emerged regarding unstable stablecoins and broader financial instability, evident in the increased prevalence of negative sentiment during this period.

\smallskip
\noindent\textbf{Post-Collapse.} Following the SVB collapse, the sentiment trend began to stabilize. Regulatory interventions and measures aimed at securing fiat money (which directly influences the stability of stablecoins) played a pivotal role in averting a more extensive crisis. While caution and uncertainty persisted, sentiment among influential authors started to show signs of recovery.

\smallskip
\noindent\textcolor{teal}{\underline{\textbf{Observation-5.\ding{206}}}: \textit{\textbf{Influencers hold greater sway in the market, and their behavioral patterns often align with the broader trends associated with \textit{\#crypto} hashtags.}}}

%===================================
\subsection{Market Behaviors}\label{subsec-market-behav}

\noindent\textcolor{teal}{\textit{\textbf{Glance.}}} To further understand the market dynamics during and post this incident, we refer to three technical indicators commonly employed in cryptocurrency trading markets: \textit{crypto fear \& greed index},  \textit{spent output profit ratio} (SOPR), and \textit{Pi cycle indicators}. The data spans a period of three months (early March - late May) and focuses on the Bitcoin topic\footnote{Data source [Aug 2023]\url{https://studio.glassnode.com/}}.

\smallskip
\noindent\textbf{Crypto Fear \& Greed index.}
This indicator is a composite value employed to gauge market sentiment in the crypto markets, with a particular focus on the Bitcoin market (though it's also applicable to the broader cryptocurrency markets). It furnishes a numerical score on a scale ranging from 0 (indicating extreme fear) to 100 (extreme greed). The sub-metrics constituting this indicator include volatility (33\%), trading volume (33\%), social media activity (20\%), market dominance (7\%), and search trends (7\%). Its primary utility lies in aiding investors in steering clear of market overreactions fueled by either extreme fear or greed.

\begin{figure}[t]
    \centering
    \subfigure[Fear \& greed index]{
    \begin{minipage}[t]{0.5\textwidth}
    \centering
    \includegraphics[width=3.6in]{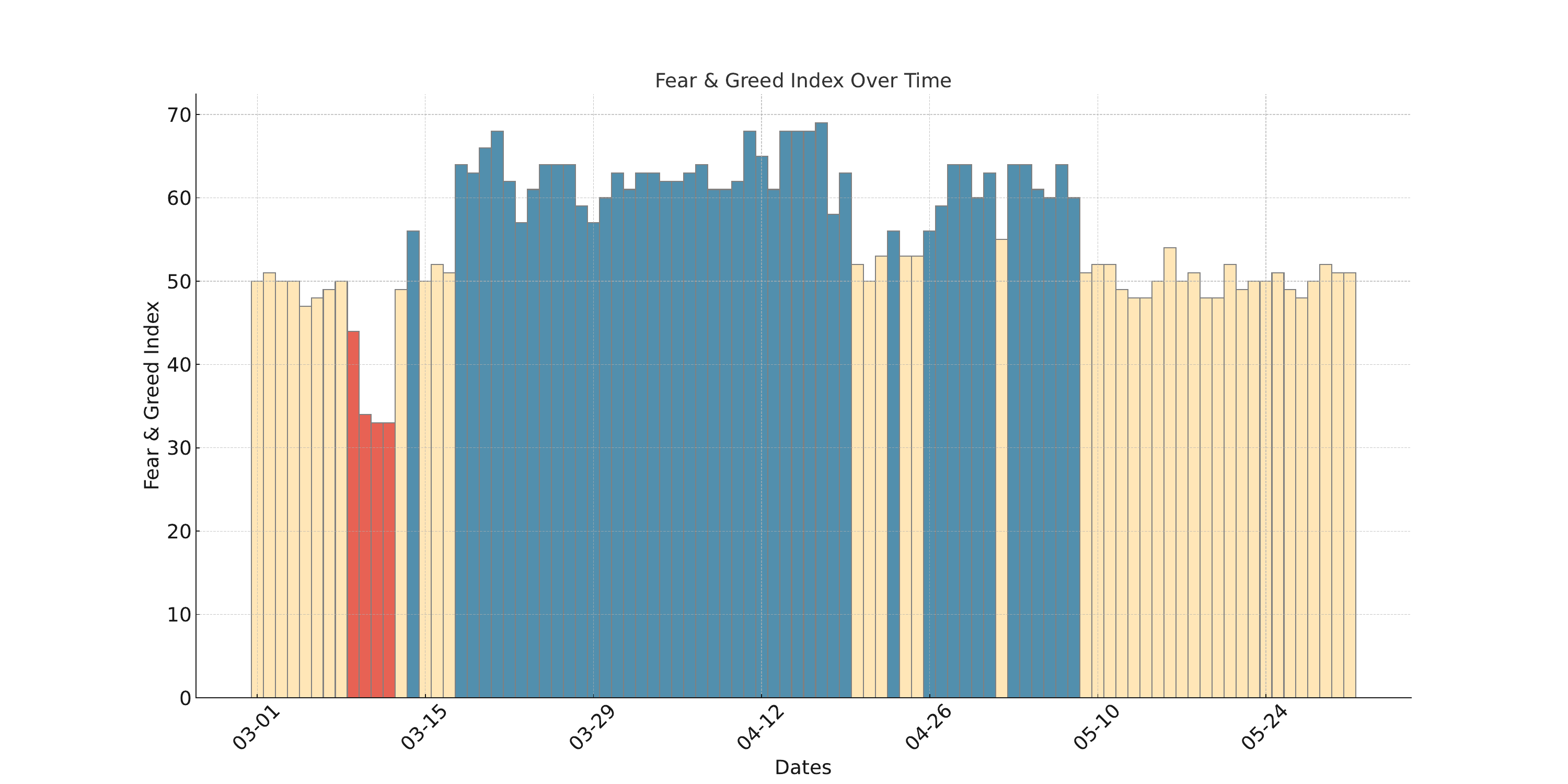}
    \end{minipage}
    \label{fig:market_behaviour_fear}
    }\vspace{-0.15in}
    \subfigure[Spent output profit ratio (SOPR)]{
    \begin{minipage}[t]{0.5\textwidth}
    \centering
    \includegraphics[width=3.6in]{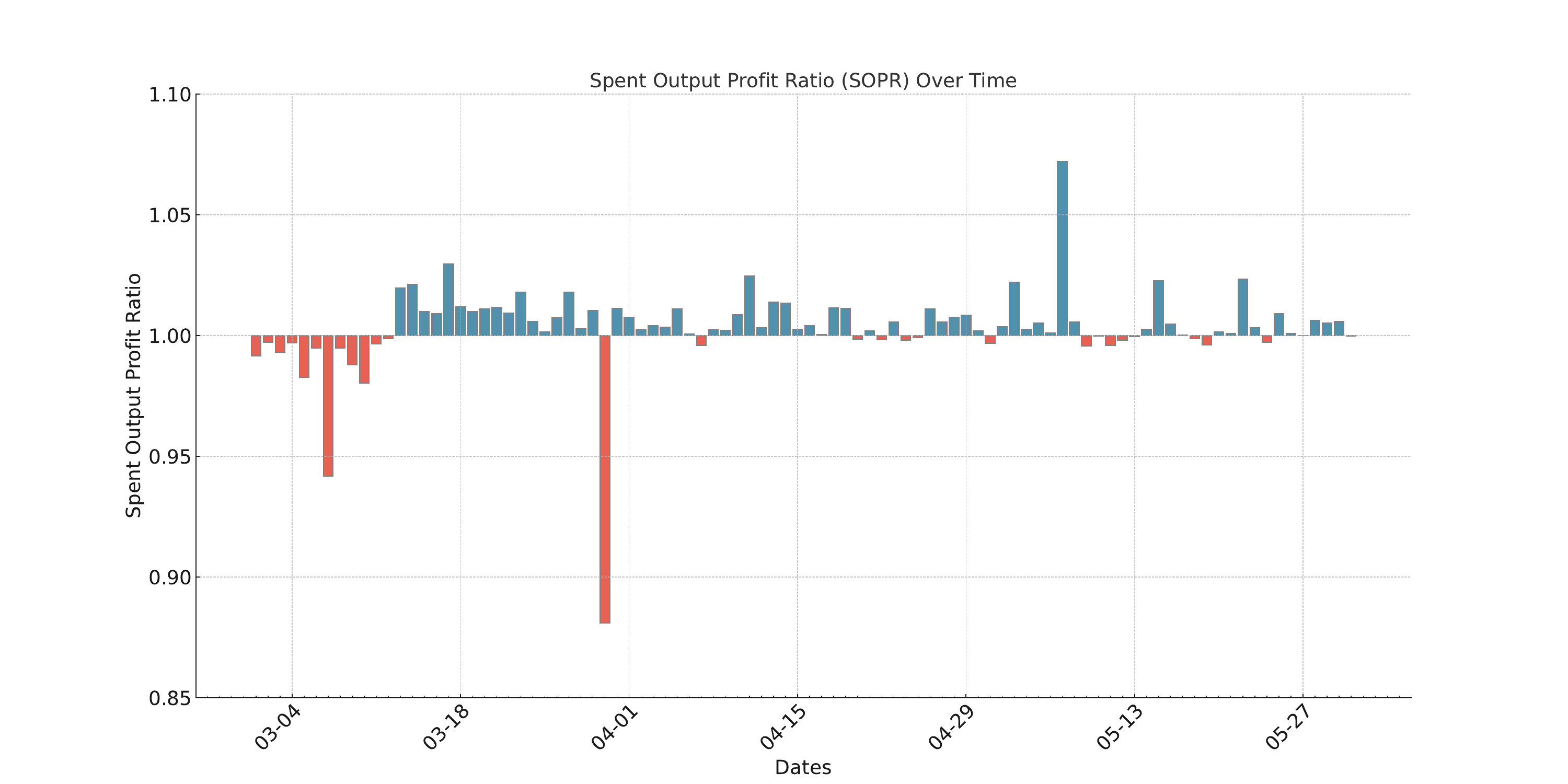}
    \end{minipage}
    \label{fig:market_behaviour_sopr}
    }\vspace{-0.15in}
    \subfigure[PI cycle indicators]{
    \begin{minipage}[t]{0.5\textwidth}
    \centering
    \includegraphics[width=3.6in]{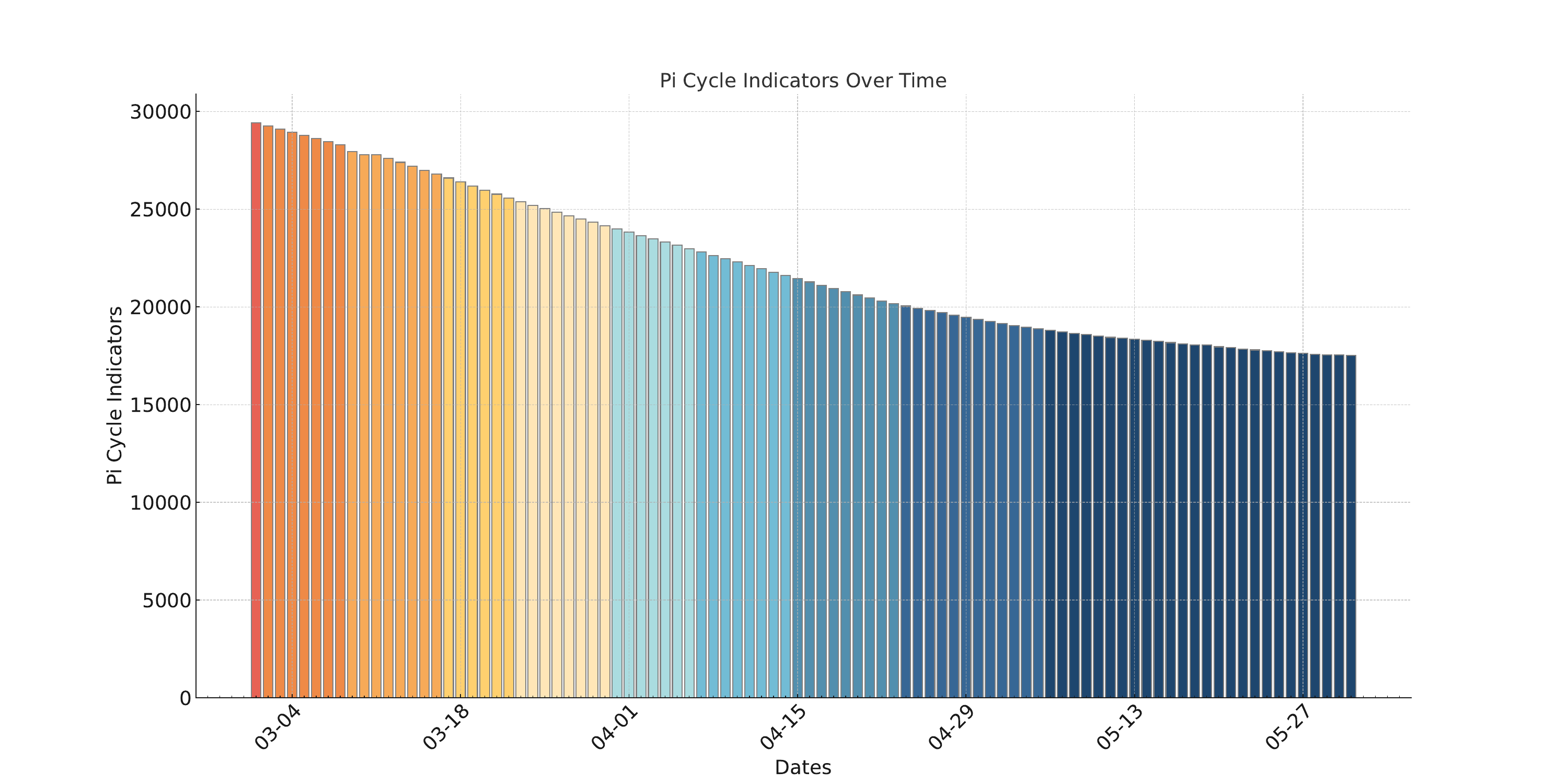}
    \end{minipage}
 \label{fig:market_behaviour_pi}
    }\vspace{-0.15in}
\caption{Market behaviour analyses}
\label{fig:market_behaviour_}
  \vspace{-0.2in}
\end{figure}

In the immediate aftermath of the SVB collapse, the crypto fear \& greed index (cf. \textcolor{teal}{Fig.\ref{fig:market_behaviour_fear}}) experienced a sharp decline, particularly notable during the period from March 9th to 13th (as indicated by the red bars). This decline was a manifestation of the heightened fear among investors. The Index's descent to the lower end of its range can be attributed to the market's sudden recognition of systemic risks, which had been unveiled by the SVB collapse. The abrupt drop in the Index indicated that investors become risk-averse, opting for safer assets rather than the more volatile cryptocurrencies. However,  the Index quickly rebounded to its previous levels (yellow bars) or even surpassed them (blue bars). This suggests that only a small subset of crypto investors exited the market while the majority retained their confidence.

\smallskip
\noindent\textbf{Spent Output Profit Ratio (SOPR).}
The indicator is a key metric widely employed in cryptocurrency trading and analysis, with a particular focus on Bitcoin. It serves as a tool for evaluating the profitability of cryptocurrencies. SOPR is computed by dividing the realized value (the price at which UTXO was initially created/spent) by the current value (the prevailing market price of the cryptocurrency). When the SOPR exceeds 1, it signifies that, on average, the tokens being spent are fetching prices higher than their acquisition, indicating a profitable scenario. Conversely, if the SOPR falls below 1, it suggests that coins are, on average, being sold at a loss, indicating a less favorable market condition.

SOPR exhibited significant volatility during this timeframe. Ordinarily, the indicator above 1 signifies a profitable market (illustrated by blue bars). However, the SVB collapse triggered a domino effect on SOPR values (see \textcolor{teal}{Fig.\ref{fig:market_behaviour_sopr}}), pushing them below the unity threshold (red bars). This suggests that investors were rapidly offloading their assets, potentially to offset losses or address urgent liquidity requirements, even if it entailed selling at a deficit. This mirrors the heightened sense of urgency and a flight-to-liquidity scenario reminiscent of the conditions SVB faced prior to its collapse. Likewise, following a brief panic period, SOPR gradually rebounded above the benchmark.

\smallskip
\noindent\textbf{Pi Cycle Indicators.}
The Pi Cycle Indicator is a tool frequently employed to pinpoint potential market peaks and troughs within Bitcoin's price cycles. It relies on two key components: the 111-day and 350-day moving averages of Bitcoin's price. When a Pi Cycle Top signal emerges (signified by the 111-day moving average crossing above the 350-day moving average), it may indicate that Bitcoin has extended itself and might be poised for a substantial correction or a bearish phase. Conversely, a Pi Cycle Bottom signal can serve as a potential entry point for investors, suggesting the conclusion of a bear market and the initiation of a new bullish cycle.

The indicators also exhibited notable fluctuations (\textcolor{teal}{Fig.\ref{fig:market_behaviour_pi}}). It became evident that the shorter moving average deviated significantly from the longer one, pointing to a cooling-off phase for Bitcoin and, by extension, broader crypto markets. However, it is worth noting that we observed a gradual narrowing of the gap between these two lines following the collapse, which signifies a warming trend. This observation aligns with our previous findings.

\smallskip
\noindent\textcolor{teal}{\underline{\textbf{Observation-5.\ding{207}}}: \textit{\textbf{Market technical indicators reveal heightened density (indicative of increased fears, reduced profitability, and substantial volatility) during a collapse, followed by a swift recovery.}}}

%=================================================  
\section{Learn from History: Related Work on Major Crypto Events to Date}
\label{sec-history}
%=================================================   

\subsection{The Wave of Tokenomics} \label{subsec-tokenwave}

The concept emerged in 2015 with intensive \textit{mining} activities based on PoW~\cite{bonneau2015sok}, where participants would use computational power to validate transactions and create new tokens. This approach laid the foundation for the decentralized nature of cryptocurrencies. The year 2017 marked the rise of \textit{initial coin offerings} (ICOs) \cite{bellavitis2021comprehensive}, which allowed projects to raise funds by exchanging Ethereum (ETH) via ERC20 \cite{erc20} for their native tokens. ICOs provided a new avenue for startups to gain monetary support from the community. \textit{initial exchange offerings} (IEOs) \cite{ieo} in 2018 enabled projects to list their tokens directly on (centralized-) cryptocurrency exchanges, allowing investors to purchase tokens using platform tokens. This streamlined the fundraising process and provided additional liquidity. Another notable development was Fcoin's \textit{transaction mining} model \cite{fcoin}, introduced in the same year. It allowed users to receive platform tokens by actively participating in trading activities to incentivize the trading volume. \textit{VDS resonance} mode \cite{vdsmode} introduced an alternative ICO distribution mechanism. Users could stake their tokens (majorly, Bitcoin) and receive rewards in return. In 2020, \textit{decentralized finance} (DeFi)~\cite{werner2022sok}\cite{jiang2023decentralized} gained significant traction. It offered financial services such as lending, borrowing, and trading in a decentralized manner. Users could provide liquidity to decentralized exchanges \cite{xu2023sok} and earn tokens as a reward, contributing to the liquidity of the ecosystem. \textit{Airdrop campaigns} \cite{allen2023airdrop} became popular as a means of distributing tokens for free. Participants could earn project tokens by just paying gas fees for transactions. In 2021, \textit{non-fungible tokens} (NFT) \cite{wang2021non} within Ethereum ecosystems gained significant attention. These unique digital assets can be used to represent ownership or proof of authenticity for all digital artworks, collectibles, and virtual real estate. The rise of \textit{play-to-earn} models \cite{yu2022sok} in 2022 revolutionized the gaming industry. Users could earn tokens by playing on-chain games to participating in Web3 space \cite{wang2022exploring}, transforming leisure time into a source of income. Most recently, in 2023, the deployment, minting, and transfer of inscriptions on Bitcoin using the \textit{BRC20 standard} \cite{binancebrc} facilitated the re-creation of unique tokens on the world's largest cryptocurrency.

\smallskip
\noindent\textcolor{teal}{\underline{\textbf{Observation-6.\ding{202}}}: \textit{\textbf{Collectively, the wave of tokenomics has shown itself to be an ever-evolving phenomenon, with changes occurring year by year. However, the fundamental nature remains constant: it consistently seeks easy (so-called efficient) and risk-free (equiv. decentralized) ways of attracting investment and distributing associated tokens.}}}

%16 年的 POW 算力产币，17 年的 ICO（ETH 换项目Token），18 年的 IEO（交易平台币换项目 Token ），Fcoin 交易挖矿（主流币换平台币），VDS 共振模式，20 年的 DeFi（提供流动性换取 Token ），刷空投（gas fee 博项目 Token ），21 年的 play to earn ...再到近期的 BRC20，在比特币上部署、铸造和转账铭文。便是一种更去中心化与更公平的 Token 分发方式。

\subsection{Human-factors in Cryptocurrency Collapse}\label{subsec-fail}

In the crypto market, three main parties are active: \textit{arbitrager} (equiv. individual traders), \textit{project launcher} (or dealer, miner, validator, exchange), and \textit{regulator} (e.g., government). Each party has a different level of power that may result in market collapses.

Arbitragers rely on their own strategies to extract profits, utilizing techniques like high-frequency trading with profit bots (frontrunning attacks \cite{zhou2021high}) or using Sniperbots \cite{cernera2022token}  to take advantage of early airdrops \cite{cernera2023ready} or NFTs \cite{wen2023nftdisk}. Some influential whales can influence market trends (a.k.a. herding behavior \cite{bouri2019herding}), while fortunately, they represent only a small portion of the overall market. 

Project launchers, on the other hand, wield considerably more power than individuals as they can employ various tactics to deceive the public. These tactics include artificially inflating the price of an asset through coordinated buying. Once the price has been pumped to a desired level, the perpetrators sell their holdings in large quantities, causing the price to plummet rapidly (commonly referred to as a \textit{pump-and-dump} \cite{xu2019anatomy}). Additionally, project launchers may intentionally abandon projects, leaving investors with worthless or significantly devalued tokens (known as a \textit{rug pull}~\cite{agarwal2023short}). In some cases, due to conflicts of interest within communities, project developers may initiate a hard fork of the chain, resulting in multiple branches operating in parallel (as exemplified by the Ethereum \textit{hard fork} \cite{hardfork}). Furthermore, profit-seeking miners may actively pursue MEV \cite{weintraub2022flash} and disrupt system stability \cite{wang2022exploring2}. Lastly, unethical project owners can lead to the crash of the entire ecosystem through fraudulent behaviors, such as the collapse of FTX  \cite{fu2022ftx,lee2023dissecting}.

The third influential force in the market comes from government entities, which can impose regulatory pressures on crypto teams. Instances include China's ICO Ban in 2017 \cite{chinaban}, leading to a sharp decline in cryptocurrency prices, followed by a regulatory crackdown in South Korea in 2018. The US Securities and Exchange Commission (SEC), in 2020, started to focus on ICOs and securities offerings \cite{secico}. Meanwhile, the Internal Revenue Service (IRS) treats cryptocurrencies as property for tax purposes. In 2021, China's Cryptocurrency Crackdown \cite{chen2022effective}, along with the announcement of stricter regulations on cryptocurrency trading, triggered another significant sell-off and clear-out (e.g., OKEX, Huobi, Gate.io) with subsequent price decline across the market. Similarly, US SEC fined Kraken \$30M for its staking-as-a-service program \cite{kraken}, and CFTC sued a crypto trader in 2022 for allegedly scamming over \$114M USD through contract swaps \cite{cftc}. Furthermore, even Ethereum~2.0 (post-\textit{Merge} \cite{merge}), known for its decentralization, faces pressure from US governance due to its OFAC-compliance validators \cite{wahrstatter2023blockchain}.

\smallskip
\noindent\textcolor{teal}{\underline{\textbf{Observation-6.\ding{203}}}: \textit{\textbf{Each party involved is rationally pursuing the optimal interests on behalf of their stands: crypto communities (both individuals and projects teams) seek full freedom of cryptoassets (driven by progressive tokenomics) but are always tangled with a high risk of losing everything. Conversely, the government assumes a restrictive role, aiming to prevent overly rapid advancements in the crypto space. However, despite its authority, the regulatory measures often struggle to keep pace with the ever-evolving developments, resulting in a persistent lag behind each successive crypto wave.}}}

%=================================================  
\section{Analyses on Collapse Impacts}
\label{sec-impac}
%================================================= 
    
\subsection{Direct Impact}

\noindent\textbf{Depressed expectation.} Despite the passage of six months since SVB's collapse, the banking sector continues to grapple with persistent uncertainty. This uncertainty is exemplified by the declining share prices of regional banks such as PacWest Bank. Many banks find themselves in situations where substantial deposit outflows have resulted in liquidity shortages, mirroring the trajectory that led to SVB's downfall. Crypto companies have strong ties with these regional business banks. The challenges faced by these banks directly affect the well-being of crypto startups. While the current situation is not yet systemic, it bears the potential to evolve in that direction, influenced by market psychology. This scenario draws striking parallels to the events leading up to the 2008 financial crisis.

\smallskip
\noindent\textbf{Difficulty in financing.} From a societal perspective, SVB's downfall is poised to compound the capital-access challenges encountered by its primary client base: tech venture capitalists and startups. The tech ecosystem heavily relies on institutions like SVB for financial support and guidance, and the disruption caused by SVB's demise could reverberate throughout the industry. For \textit{startups}, the consequences could be dire. The inability to meet payroll obligations could lead to an exodus of key talent, as employees may seek more stable employment opportunities elsewhere or, in the worst-case scenario, face layoffs. In addition to startups, \textit{banks} with substantial bond holdings are now under heightened scrutiny from investors. They face increased pressure to align their funding costs with the rates earned on their assets, potentially exposing vulnerabilities in their balance sheets. Moreover, there is a growing challenge posed by tightening online \textit{regulations}. These regulations are not only making it harder for tech firms to generate revenue but also increasing the costs associated with compliance. This dual impact can strain the profitability and sustainability of tech enterprises, affecting their ability to contribute to economic growth and innovation. Impacts on the society side will directly decrease the investment in the crypto market and Web3 projects.

\smallskip
\noindent\textbf{More powerful government.} Government intervention has played a pivotal role in addressing this crisis, with swift reactions and assertive actions, however, leading to an expansion of government authority. The impact of these measures is evident across various facets of the market, including the management of interest rates. The Federal Reserve's decision to implement rate hikes as a means to combat inflation has significantly influenced depositor behavior, driving them to seek higher-yielding options. This shift has exacerbated liquidity challenges faced by banks such as Silvergate and Signature Bank, posing operational difficulties. Interestingly, much like the rate hikes that preceded the 2008 financial crisis, the intended solution may inadvertently contribute to worsening the situation. Another key aspect is the utilization of government policies. Governments can introduce a range of emergency measures to exercise macroeconomic control over the entire financial market. This is done to prevent market overreactions from escalating liquidity issues into solvency crises. Institutions like Credit Suisse, which have had to rely on central banks as lenders of last resort, serve as a stark reminder of the potential escalation of the crisis. In times of crisis, the entire market is at risk of collapsing, including the crypto markets.

\subsection{Compared to 2008}\label{subsec-compare2018}

Several factors contribute to the observation that the impact of the SVB event in 2023 is not as significant as the 2008 financial crisis~\cite{spiegel2011academic}, as commonly recognized:

\smallskip
\noindent\textcolor{teal}{\ding{49}} \textbf{\textit{Systemic factors.}} The 2008 financial crisis stemmed from the collapse of the housing market, affecting a broad spectrum of individuals and institutions worldwide. In contrast, SVB's collapse in 2023 has a distinct cause and is less systemic in nature, primarily affecting its customer base, which comprises a relatively small group of crypto companies within the broader crypto market.

\smallskip
\noindent\textcolor{teal}{\ding{49}} \textbf{\textit{Role of cryptocurrency.}} Unlike in 2008, cryptocurrencies like Bitcoin now play a significant role in the financial ecosystem. They are often seen as alternative investments, especially during financial turmoil, serving as a hedge against traditional market movements. The stronger underlying economy in 2023 provides a different context for the cryptoasset market.

\smallskip
\noindent\textcolor{teal}{\ding{49}} \textbf{\textit{Market resilience.}} Over time, financial markets and institutions have enhanced their resilience and improved risk management practices. These developments contribute to mitigating the impact of economic shocks. Moreover, the banking sector's reduced leverage and holdings of safer assets like U.S. Treasuries have bolstered their ability to receive timely liquidity support from institutions such as the Federal Reserve, further enhancing market resilience.

\smallskip
\noindent\textcolor{teal}{\ding{49}} \textbf{\textit{Regulatory changes.}} Substantial regulatory reforms have been implemented since the 2008 crisis to fortify the stability of the financial system. These regulations have the potential to mitigate the impact of events similar to SVB's collapse. Governments and central banks have also taken proactive measures to stabilize the economy and financial markets, thereby limiting the extent of potential repercussions.

\subsection{Any Further Crypto Crisis?}

We first provide our answer based on previous data-centric analyses and qualitative investigations: \textit{It is unlikely that the crypto market will experience another worse crisis caused by SVB's influence. However, due to macroeconomic depressed expectations, the crypto market may not rapidly regain its peak levels as seen in 2020.}

Cryptocurrencies in 2023 are often considered alternative investments, especially in times of financial stress, as a hedge against traditional market movements. This behavior could be observed in how investors flocked to stable assets during the SVB Financial Group crisis, similar to how they did in the post-pandemic period of 2020 and 2021. Additionally, PayPal recently announced plans to offer the services to exchange cryptoassets with RWA assets within their apps. This is expected to greatly facilitate the user base of cryptocurrencies. As a result of these developments, cryptocurrencies are evolving into a vital component of the real-world financial infrastructure (\textit{\textbf{role of cryptocurrency}}) rather than remaining in the realm of parallel \textit{shadow} assets.

From \textcolor{teal}{Sec.\ref{subsec-compare2018}}, we can also observe that the underlying economy is fundamentally stronger now (even suffering COVID-19), acting as a buffer against systemic financial shocks and thereby reducing the risk of a financial crisis (\textit{\textbf{market resilience}}). Enhanced regulatory measures (\textit{\textbf{regulatory changes}}) and more vigilant Federal Reserve oversight contribute to a system where individual bank failures are less likely to metastasize into larger crises. As a subset of the overall financial system, crypto crises follow global trends.

Moreover, the current lower levels of leverage and the nature of investments in safer assets like U.S. Treasuries further reinforce the robustness of the financial sector (NOT \textit{\textbf{systemic factor}}). In particular, conventional banks may become more conservative, leading to slower consumer spending and potentially reduced inflation. This can help mitigate the potential for financial bubbles, as demonstrated by counterexamples~\cite{fu2023rational,fu2022ftx}. While liquidity issues can still pose a threat, as demonstrated by the failure of SVB, these are more manageable due to the Federal Reserve's policies and tools for intervening in such situations.

%==============================================
\section{User Perception}
\label{sec-userpcpt}
%==============================================

\subsection{Coexist: Pessimism \& Optimism}\label{subsec-user-coexist}

Based on our investigations in \textcolor{teal}{Sec.\ref{sec-emprical}}, we observed that crypto users displayed a spectrum of sentiments following the collapse of SVB, ranging from pessimistic to optimistic.

\smallskip
The first type tends to lean towards negativity, with most observations supported by the content of their tweets and actual price drops in the crypto markets (\textcolor{teal}{\textbf{Observation-5.\ding{202}}}\&\textcolor{teal}{\textbf{5.\ding{203}}}).

\begin{itemize}
    \item \textit{Distrust in finance.} The collapse fueled distrust in traditional financial institutions, exposing vulnerabilities and inadequacies even within established banks. This sentiment also led some users to question the decentralization and security of cryptocurrencies.

    \item \textit{Loss aversion}. Some users experienced a sense of loss aversion, as the SVB collapse served as a reminder of the potential risks associated with the crypto market. They may have become more risk-averse and considered diversifying their investments or reducing exposure to crypto assets.

    \item \textit{Increased skepticism.} Skepticism increased among some crypto users, particularly towards centralized platforms and institutions (centralized crypto exchanges included). They became more discerning when choosing where to store their crypto assets and transact.
\end{itemize}

The second type tends to exhibit a neutral sentiment. However, many crypto users with this sentiment may choose not to express their views publicly or opt for inaction (\textcolor{teal}{\textbf{Observation-5.\ding{204}}}\&\textcolor{teal}{\textbf{5.\ding{206}}}). 

\begin{itemize}
    \item \textit{Concern and caution.} Many crypto users expressed concerns about the stability of financial institutions, including those within the cryptocurrency industry. They became cautious and reevaluated the safety of their crypto assets, emphasizing the importance of secure storage and risk management.
    
    \item \textit{Silence.} In our opinion, we think a very large portion of crypto users might have chosen to remain silent and refrain from expressing their opinions or sentiments publicly. This group may have preferred to observe developments quietly or not engage in discussions about the event. However, this category is not readily observable.

    \item \textit{Indifference or do not care.} Some crypto users may have reacted with indifference or apathy to the SVB collapse. They might not have seen the event as directly impacting their investments or their perception of the cryptocurrency market.
\end{itemize}
    
The third type tends to exhibit resilience and positivity, which is primarily reflected in the swift recovery (\textcolor{teal}{\textbf{Observation-5.\ding{205}}}\&\textcolor{teal}{\textbf{5.\ding{207}}}) of token prices and crypto-related activities (compared to levels just before the collapse rather than its peak time in 2022).

\begin{itemize}
    \item \textit{Long-term optimism.} 
     Despite the SVB collapse, some crypto users remained resilient and maintained a long-term optimistic outlook on the cryptocurrency market. They viewed the event as a learning experience and an opportunity for the industry to improve its risk management practices.
    
    \item \textit{Calls for regulation.} In response to the SVB collapse and potential systemic risks, some crypto users advocated for increased regulatory oversight in the crypto space. They believed that regulation could provide a level of protection and accountability.

    \item \textit{Advocacy for decentralization.} The collapse reinforced the belief in the decentralization principles of cryptocurrencies for some users. They saw decentralization as a means to reduce reliance on traditional financial institutions and mitigate systemic risks.
    
\end{itemize}

\subsection{Lessons for Crypto Users}\label{subsec-user-lesson}

The lessons are general but highly relevant for crypto traders or participants due to the higher liquidity, leverage, and risk compared to traditional traders in stocks or options.

\smallskip
\noindent\textbf{Do not miss due diligence research.}
The collapse of SVB serves as a reminder of the crucial role that due diligence and independent research play in investment decisions. Prior to its collapse, SVB had often received high ratings and was considered stable, yet underlying issues in its balance sheets went unnoticed. The event compels investors to engage in more thorough analyses of financial institutions where they store their digital assets, even those that have previously enjoyed a strong reputation.  In the context of cryptocurrencies, users must exercise heightened caution, as many crypto tokens lack substantial reserves and the industry operates with limited regulation. This environment can create opportunities for crypto projects to operate without significant fear of consequences. Real-word examples can refer to crypto frauds such as rug pull~\cite{jiang2023decentralized}, pump-and-Dump~\cite{xu2019anatomy}, Sniper bots~\cite{cernera2023token} and ICO/IEO/IDO (\textcolor{teal}{Sec.\ref{subsec-tokenwave}}).

\smallskip
\noindent\textbf{Asset diversification strategy.}
The SVB collapse exacerbated the existing volatility within the crypto market, shining a spotlight on the perils of unrealized threats hidden within balance sheets. One significant factor contributing to the collapse was an inflexible investment strategy, primarily reliant on long-term fixed bonds and securities. These investments resulted in substantial unrealized losses in the bank's balance sheet, which unfortunately went unnoticed by the risk control department. This not only led to the bank's collapse but also resulted in the loss of customers' principal funds, triggering a ripple effect across the entire crypto market. The consequences underscore the critical need for flexible asset diversification strategies. Prior to this event, many investors had either heavily leaned toward traditional finance or had become overly invested in cryptocurrencies, in addition to often relying on a single strategy without incorporating any hedges for protection. Such a singular strategy created inflexibility and elevated the risk of being caught off-guard. The aftermath reignites the emphasis on diversifying across multiple strategies to mitigate the risks associated with systemic shocks.

\smallskip
\noindent\textbf{Understanding hidden connectedness.}
It's crucial to understand the interconnectedness of today's financial landscape, where a downfall in one sector can easily spill over into another. The collapse of traditional financial institutions like SVB vividly illustrates how crises in the traditional banking sector can rapidly impact the cryptocurrency market. Despite SVB being a regional small bank, its collapse had far-reaching consequences for the entire crypto industry. This serves as a stark reminder that crypto investors should have a comprehensive understanding of the entire life cycle of how a crypto token creates value. This understanding should extend beyond smart contract token creation and exchange on decentralized exchanges (DEXes) to include a clear comprehension of where their RWAs are allocated and how those RWAs are expected to circulate within the crypto ecosystem.

\vspace{-0.1in}

%==============================================
\section{New challenges?}
\label{sec-chall}
%==============================================

\noindent\textbf{Governance dilemmas.}
The collapse of SVB serves as an unfortunate yet instructive case study of governance, which applies not only to traditional financial institutions but also to crypto companies. Governance encompasses both internal cooperative governance and external societal governance, often influenced by government regulations.
Our investigation reveals that a core issue of collapse lies in internal governance, particularly in risk management and internal controls. The bank operated without a chief risk officer for an extended period, despite warnings from federal regulators about the inadequacy of its risk control mechanisms. Many crypto companies adopt a flat management structure, potentially exposing them to greater risks due to the \textit{diffusion of responsibility}~\cite{darley1968bystander}.
Externally, the recent Federal Reserve rate shifts have triggered events leading to the collapse. Governments employ various policies that can influence the entire financial system~\cite{smith1937wealth}, and crypto markets are not immune to these effects. This has led to a degree of centralization in the crypto space, resulting in increased regulations. The U.S. SEC, for instance, has initiated numerous actions and censorship measures (e.g., OFAC~\cite{ofac}) against crypto companies~\cite{secico,tornadocashsanction,wang2023blockchain}, raising concerns. While some crypto enthusiasts advocate for complete decentralization~\cite{aftermath}, this also carries the risk of disorder and challenging decision-making processes.

\smallskip
\noindent\textbf{Algorithmic design pitfalls.}
SVB also provides financial services to crypto companies and technological support, particularly algorithms, to optimize resource allocation. These algorithms can be used across diverse domains, including trading, risk assessment, and the implementation of automated investment strategies. Financial institutions, including SVB, have the capacity to broaden their support by providing essential tools such as Application Programming Interfaces (APIs) and oracles~\cite{pasdar2023connect}, which serve as data conduits to crypto platforms.
However, the concern about the effectiveness in mitigating SVB's internal risks while simultaneously safeguarding traders/investors from vulnerabilities still exists.  Equally important is the question of algorithmic transparency: \textit{Did these algorithms operate in a transparent manner, providing comprehensive insight into their decision-making processes?} Alternatively, \textit{did they function as ``black boxes", obscuring the rationale behind their actions?} If SVB's risk management algorithms are found deficient, the potential ramifications may extend into the crypto space, influencing traders to make decisions that are inadequately informed.

\smallskip
\noindent\textbf{Barrier in the merging of CeFi and DeFi.} After the collapse of SVB, which was also impacted by a series of other collapses such as Three Arrows~\cite{threearrow}, Luna~\cite{fu2023rational}, and FTX~\cite{fu2022ftx}, many crypto investors have shifted their behavior. They have become cautious about depositing their cryptoassets into traditional financial services due to concerns and a loss of confidence in singular organizations. Instead, they are opting to deposit their funds into crypto wallets and entrust their private keys. The recurring failures of centralized giants have significantly eroded investor confidence and dampened their enthusiasm for CeFi services. This crisis has also accelerated the transition from the Web2 to the Web3 framework~\cite{yu2023web3}, as decentralized platforms gain prominence in an environment where centralized institutions have shown vulnerabilities. Furthermore, the permeation of apprehensions from the traditional financial industry has created a barrier for DeFi communities. As crypto assets become increasingly integrated with mainstream finance, the crypto market is no longer insulated from shocks in the traditional financial sector. For example, if traditional banks adopt more conservative practices and begin hoarding cash, as hinted in earlier discussions, it could also reduce liquidity in the crypto market. Similarly, regulatory hurdles pose concerns. In 2023, the likelihood of increased regulation is on the rise, mirroring the conservative lending practices. This has led to discussions enabling Web3-based Environmental, Social, and Governance (ESG) initiatives as a way to enhance the ESG profiles of crypto markets~\cite{9805516}, drawing important lessons from the SVB collapse.

%=================================================  
\section{Conclusion}
\label{sec-conclu}
%=================================================   
The SVB collapse reverberated across the entire financial markets, leaving no stone unturned. Through our comprehensive study that covers design exploration, factual summation, sentiment analyses, and market performance, we demonstrate the resilience of cryptocurrencies and their coherence with macroeconomics. In our view, this study explores the impact of the SVB collapse on the crypto space with both \textit{completeness} and \textit{depth}, filling the absence.

%=================================================
%=================================================
\bibliographystyle{unsrt}
\bibliography{bib.bib}
%=================================================
%=================================================

\appendix

\section{Additional Finance Basics}
\label{appn-finance}

\noindent\textbf{Bond.}  A bond is a fixed-income investment instrument where an investor loans money to companies or governments for a specific period, with the promise of periodic interest payments (or coupon payments) and the return of the principal amount at the bond's maturity date.  Bonds are considered to be lower-risk investments, as they offer a fixed income stream and are predictable. 

\smallskip
\noindent\textbf{Government debt.} Government debt, also known as public debt or national debt, refers to the total amount of money that a government owes to creditors. Governments borrow money by issuing debt securities (e.g., \textit{\textbf{government bonds}} or Treasury bills) to investors. Government debt is a critical aspect of fiscal policy, ensuring the sustainability of a nation's economy.

\smallskip
\noindent\textbf{Interest rate}. It represents a percentage charged on the principal amount of a loan or an investment, serving as a measure of the cost of borrowing money. Interest rates can vary due to various factors and we present its correlation to those factors in \textcolor{teal}{Tab.\ref{tab-correlation}}.

\begin{wraptable}{r}{4cm}
    \vspace{-0.15in}
    \caption{Interest rate vs ...} \label{tab-correlation}
    \vspace{-0.15in}
    \resizebox{\linewidth}{!}{
    \begin{tabular}{c|c} 
    \cellcolor{gray!10}\textbf{\textit{Bond}}  &  \cellcolor{gray!10}negative  \\
    \cellcolor{gray!10}\textbf{\textit{Inflation rate}} & \cellcolor{gray!10} negative  \\
    \cellcolor{gray!10}\textbf{\textit{Stocks}} & \cellcolor{gray!10} weak negative  \\
    \cellcolor{gray!10}\textbf{\textit{Gold}} &  \cellcolor{gray!10}uncertain  \\
    \cellcolor{gray!10}\textbf{\textit{Housing}} & \cellcolor{gray!10} uncertain  \\
    \cellcolor{gray!10}\textbf{\textit{Currency}} & \cellcolor{gray!10} positive  \\
    \cline{2-2}
    \multicolumn{1}{c}{\quad\textbf{}\quad}   & \multicolumn{1}{c}{\cellcolor{gray!10}\quad\textit{\textbf{Correlation}}\quad}  \\
    \end{tabular}
    }
    \vspace{-0.15in}
\end{wraptable}

\smallskip
\noindent\textbf{Yield curve}. The curve represents the relationship between the interest rates ($y$ axis) and the time to maturity of bonds ($x$ axis). A normal yield curve is upward-sloping, indicating that longer-term bonds have higher yields than shorter-term bonds (due to the increased risk/uncertainty over a long period). In contrast, an inverted yield curve (short-term rates higher than long-term rates) can be seen as a predictor of an economic recession.

\smallskip
\noindent\textbf{10-year benchmark}. Typically, this notion refers to the yield on the \textbf{\textit{10-year U.S. Treasury note}}, which serves as a benchmark for measuring interest rates across the economy. As the U.S. government is considered to have a very low risk of default, Treasury yields are viewed as a \textbf{\textit{risk-free (interest) rate}} of return and used as a benchmark for other investments. A rise in this yield may indicate investors' optimism about economic growth and inflation, while a fall may signal concerns about economic weakness or deflation.

\section{Financial Services and DeFi}
\label{appn-defi}

\noindent\textbf{Financial services in CeFi.} Centralized Finance is the traditional financial system where all services are provided by intermediaries. Such reliable third parties can facilitate financial activities (e.g., deposits, withdrawals, lending). CeFi offers products including:

\begin{itemize}
    \item \textit{Stocks}: a stake representing a share of ownership of a company and constituting a claim on part of company’s earnings.
    \item \textit{Bonds}: as described in Appendix~\ref{appn-finance}. 
    \item \textit{Options}: financial contract linked to the value of underlying securities, encompassing stocks, bonds, commodities, and indexes. It grants holders the right to buy or sell these assets at a predetermined price and within a specified timeframe.
    \item \textit{Mutual funds}: investment strategies that allow individuals to pool their money together with other investors to purchase a collection of stocks, bonds, or other securities.
    \item \textit{Exchange-Traded Funds (ETFs)}: similar to mutual funds but traded on stock exchanges, designed to track the performance of a specific index, sector, commodity, or asset class.
    \item \textit{Futures Contracts}: agreements to buy or sell an asset at a predetermined price on a future date.
    \item \textit{Foreign Exchange (Forex) market}: the global market for trading currencies, where participants exchange one currency for another at an agreed-upon exchange rate.
    \item \textit{Real estate investment trusts (REITs)}: A certificate similar to stocks and ETFs and traded on major stock exchanges, used to invest in real estate without owning physical properties.
    \item \textit{Certificates of deposit (CDs)}: Time deposits offered by banks with fixed terms and interest rates (predictable returns).
    \item  \textit{Annuities}: Insurance contracts that provide regular payments to the annuitant, typically used for retirement income.

\end{itemize}

\smallskip
\noindent\textbf{Mapping to DeFi.} DeFi protocols are financial services built on blockchain technology that eliminates intermediaries via the use of smart contracts. DeFi products have covered a wide range of functions in CeFi services, including DEXs (Uniswap, dYdX), lending (Compound, Aave), yield aggregators (Convex), staking (Lido), and others. We compare them in \textcolor{teal}{Tab.\ref{tab:mapping}}.

\textit{Uniswap} is a protocol on Ethereum for swapping ERC20 tokens without the need for buyers and sellers to create demand. \textit{Curve Finance} is a decentralized exchange optimized for stablecoin trading providing low-risk, low-slippage trades. \textit{Aave} is a decentralized lending system that allows users to lend, borrow, and earn interest on crypto assets. \textit{MakerDAO} is a decentralized credit platform on Ethereum that supports Dai, a stablecoin whose value is pegged to USD. \textit{Compound} is a protocol on the Ethereum blockchain that establishes money markets, which are pools of assets with algorithmically derived interest rates, based on asset supply and demand.

\vspace{-0.1in}
\begin{table}[!hbt]
\centering
\caption{Mappings between CeFi and DeFi}
\label{tab:mapping}
\vspace{-0.15in}
\resizebox{\linewidth}{!}{
\begin{tabular}{c|cc}

 \multicolumn{1}{c}{\textbf{\textit{CeFi}}} & \multicolumn{1}{c}{\cellcolor{gray!10}\textbf{\textit{DeFi}}} & \multicolumn{1}{c}{\cellcolor{gray!10}\textbf{\textit{Products}}} \\ 
 \cline{2-3}
\cellcolor{gray!10}\textit{Stocks} & \cellcolor{gray!10} ERC20-style tokens & \cellcolor{gray!10} ICOs [by smart contract]\\ 
\cellcolor{gray!10}\textit{Bonds} & \cellcolor{gray!10} Algorithmic Stablecoins & UST(-Luna)/Ampleforth \cellcolor{gray!10}  NFT \\ 
\cellcolor{gray!10}\textit{Mutual funds} & \cellcolor{gray!10} Crypto Fund & \cellcolor{gray!10} -  \\ 
\cellcolor{gray!10}\textit{ETFs} & \cellcolor{gray!10} Crypto Index  & \cellcolor{gray!10} Fear\&Greedy Index\\ 
\cellcolor{gray!10}\textit{Futures} & \cellcolor{gray!10} Futures in CEX/DEX & \cellcolor{gray!10} [by order-book/SC]  \\ 
\cellcolor{gray!10}\textit{Forex} & \cellcolor{gray!10} Native tokens & \cellcolor{gray!10} [cross-chain tech] \\ 
\cellcolor{gray!10}\textit{REITs} & \cellcolor{gray!10} NFTs & \cellcolor{gray!10}Pepe [ERC-20/BRC-20]  \\ 
\cellcolor{gray!10}\textit{CDs} & \cellcolor{gray!10} Staking  & \cellcolor{gray!10} Ethereum2.0 [PoS] \\ 
\cellcolor{gray!10}\textit{Annuities} & \cellcolor{gray!10} N/A   & \cellcolor{gray!10} - \\ 
\cellcolor{gray!10}\textit{Exchanges} & \cellcolor{gray!10} EX/DEX  &  \cellcolor{gray!10} Binance/Uniswap \\ 
\end{tabular}
}
\end{table}
\vspace{-0.05in}

\smallskip
\noindent\textbf{Relation.} 
CeFi and DeFi have a symbiotic relationship and together form the broader financial market. The key distinction between CeFi and DeFi lies in their operations and services. While CeFi requires intermediaries and trust, DeFi operates on smart contracts and transparency. The following comparison (\textcolor{teal}{Tab.\ref{tab:CeFiVsDeFi}}) summarizes the differences between CeFi and DeFi. The missing points in the CeFi market, such as lack of transparency, inefficiencies, and exclusivity, are addressed by DeFi. However, the DeFi market is still maturing and presents its own set of challenges, including smart contract vulnerabilities and high price volatility. 

\vspace{-0.05in}
\begin{table}[!hbt]
\centering
\caption{Comparison between CeFi and DeFi}
\label{tab:CeFiVsDeFi}
\vspace{-0.15in}
\resizebox{\linewidth}{!}{
\begin{tabular}{c|cc}

 \multicolumn{1}{c}{} & \multicolumn{1}{c}{\cellcolor{gray!10}\textbf{\textit{CeFi}}} & \multicolumn{1}{c}{\cellcolor{gray!10}\textbf{\textit{DeFi}}} \\ \cline{2-3}
\cellcolor{gray!10}\textbf{\textit{Control}} & \cellcolor{gray!10} By institutions & \cellcolor{gray!10}By automated code (SC) \\ 
\cellcolor{gray!10}\textbf{\textit{Access}} & \cellcolor{gray!10} Permissioned (KYC/AML) & \cellcolor{gray!10}\cellcolor{gray!10}Permissionless \\ 
\cellcolor{gray!10}\textbf{\textit{Transparency}} & \cellcolor{gray!10}Low & \cellcolor{gray!10}High  \\ 
\cellcolor{gray!10}\textbf{\textit{Custody}} & \cellcolor{gray!10}By third parties & \cellcolor{gray!10}Self-custody \\ 
\cellcolor{gray!10}\textbf{\textit{Intermediary}} & \cellcolor{gray!10} TTP/Authorities & \cellcolor{gray!10} None \\ 
\end{tabular}
}
\end{table}

%=================================================
%=================================================

\end{document}